\documentclass[lineno]{jfm}

\usepackage{graphicx}
\usepackage{caption}
\usepackage{subcaption}
\usepackage{subcaption}
\usepackage{newtxtext}
\usepackage{newtxmath}
\usepackage{natbib}
\usepackage{hyperref}
\usepackage[normalem]{ulem}

\hypersetup{
    colorlinks = true,
    urlcolor   = blue,
    citecolor  = black,
}

\newcommand{\RomanNumeralCaps}[1]
\linenumbers

\usepackage{cleveref}

\crefformat{section}{\S#2#1#3}

 \shortauthor{N. Singh and A. Pal}
 \shorttitle{Turbulent mixing driven by the Faraday instability in rotating miscible fluids}
\title{Direct numerical simulations of turbulent mixing driven by the Faraday instability in rotating miscible fluids.}
\author{Narinder Singh\aff{1}
 and Anikesh Pal\aff{1}
  \corresp{\email{pala@iitk.ac.in}}}

\affiliation{\aff{1}Department of Mechanical Engineering, Indian Institute of Technology Kanpur, Kanpur 208016, U.P., India}

\begin{document}
\maketitle

\begin{abstract}
The effect of the rotation on the turbulent mixing of two miscible fluids of small contrasting density, induced by Faraday instability, is investigated using direct numerical simulations (DNS). We quantify the irreversible mixing which depicts the conversion of the available potential energy (APE) to the background potential energy (BPE) through irreversible mixing rate $\mathcal{M}$. We demonstrate that at lower forcing amplitudes, the turbulent kinetic energy ($t.k.e.$) increases with an increase in the Coriolis frequency $f$ till $\left(f/\omega\right)^2<0.25$, where $\omega$ is the forcing frequency, during the sub-harmonic instability phase. This enhancement of $t.k.e.$ is attributed to the excitement of more unstable modes. The irreversible mixing sustains for an extended period with increasing $\left(f/\omega\right)^2$ till $0.25$ owing to the prolonged sub-harmonic instability phase and eventually ceases with instability saturation. When $\left(f/\omega\right)^2 > 0.25$, the Coriolis force significantly delays the onset of the sub-harmonic instabilities. The strong rotational effects result in lower turbulence because the bulk of the APE expends to BPE, decreasing APE that converts back to $t.k.e.$ reservoir for $\left(f/\omega\right)^2 > 0.25$. Therefore, in the subsequent oscillation, the  $t.k.e.$ available to contribute to the external energy input from periodic forcing is small. Since the instability never saturates for $\left(f/\omega\right)^2 > 0.25$, conversion of APE to BPE via $\mathcal{M}$ continues, and we find prolonged irreversible mixing. At higher forcing amplitudes, the instability delaying effect of rotation is negligible, and the turbulence is less intense and short-lived. Therefore, the irreversible mixing phenomenon also ends quickly for $\left(f/\omega\right)^2<0.25$. However, when $\left(f/\omega\right)^2>0.25$, a continuous irreversible mixing is observed. We also examine the mixing efficiency in terms of $\mathcal{M}$ and find that the mixing is efficient at lower forcing amplitudes and rotation rates of $\left(f/\omega\right)^2 > 0.25$ because the major portion of APE expends to BPE. \\
\end{abstract}

  
%{\bf MSC Codes }  {\it(Optional)} Please enter your MSC Codes here

\section{Introduction} \label{sec:intro} 
The Faraday instability \citep{faraday1831xvii} describes the generation of standing wave patterns at the interface of two immiscible fluids subjected to vertical periodic vibration. However, in a stably stratified two-layer miscible fluid system, these standing waves become highly disorganized above a certain forcing amplitude and start to interact with each other resulting in the mixing of fluids which eventually saturates \citep{zoueshtiagh2009experimental,amiroudine2012mixing,diwakar2015faraday}. \cite{grea2018final,briard2019harmonic} studied the onset and saturation of turbulent mixing owing to Faraday instability in fluids of small contrasting density for a wide range of parameters using theoretical models and DNS. They demonstrated that the diffuse interface begins to oscillate with small amplitude in the harmonic instability phase, followed by the rapid growth and development of a turbulent mixing zone in the sub-harmonic instability phase. However, owing to the saturation of the instability, the turbulent mixing cannot be sustained for long. They also estimated the final size of the mixing zone analytically as follows: 
\begin{equation}
\label{Lsat}
L_{sat}=\frac{2\mathcal{A}g_0}{\omega^2}\left(2F+4\right),
\end{equation}
where $L_{sat}$ is the final asymptotic size of the mixing zone (defined later) in saturation state, $\mathcal{A}=\left(\rho_1-\rho_2\right)/\left(\rho_1+\rho_2\right)$ is the Atwood number expressing the density contrast between heavy fluid ($\rho_1$) and light fluid ($\rho_2$), $g_0$ is the mean acceleration, $F$ is the oscillation amplitude and $\omega$ is the forcing frequency. \citet{briard2019harmonic} numerically presented the quantification of irreversible mixing associated with turbulent mixing by partitioning the total potential energy (PE) into background potential energy (BPE) and available potential energy (APE). They reported that the BPE signifying the measure of irreversible mixing increases during the onset of sub-harmonic instability. The APE, which represents the fraction of the total PE that can be transferred to BPE through irreversible mixing, peaks at saturation and is partially released in the flow as BPE. This results in more irreversible mixing, which causes the numerically computed final mixing zone size $L$ to exceed the analytically predicted $L_{sat}$ \ref{Lsat}. \cite{briard2020turbulent} performed a linear stability analysis accounting for the vertical inhomogeneity of the background density profile and viscous dissipation to predict the onset of the Faraday instability, either from a sharp or a diffuse interface and its eventual saturated state. Further experiments were carried out by \cite{briard2020turbulent,cavelier2022subcritical} using fresh and salty water to explore the dynamics of turbulent mixing, and DNS and theoretical predictions confirmed their findings. They found that when the instability is triggered, a natural wavelength appears at the interface between the two fluids. With the increase in amplitude, well-defined structures form that break due to the destabilization process at the nodes to produce turbulent mixing. Turbulence is eventually inhibited by instability saturation, and the final size of the mixing layer is consistent with the analytically predicted $L_{sat}$. %However, owing to the saturation of the instability, the turbulent mixing cannot be sustained for long. 
Recently, \citet{singh2022onset} presented a theoretical analysis of Faraday instability in miscible fluids under the effect of rotation and corroborated their findings with DNS. They reported that rotational effects %stabilize the flow and
delay the onset of the sub-harmonic instability at lower forcing amplitudes ($F$), owing to the presence of a stable region between the harmonic to sub-harmonic transition and before the onset of the sub-harmonic instabilities. In the stable region, the mixing layer grows due to diffusion. However, with an increase in $F$, the %stabilizing
instability delaying effect of rotation diminishes due to the increasing area of all the unstable regions in the 3D stability diagram, resulting in the majority of excited modes being unstable. They concluded that the instability saturates for $\left(f/\omega \right)^2<0.25$ and the mixing zone size asymptotes. However, for $\left(f/\omega \right)^2\geq0.25$, the instability does not saturate, and the mixing zone size continues to grow.  \\

%{\color{red} The following paragraph is fine, but it appears suddenly. There is no direct connection between the previous paragraph and this one. You should think about how to fill this discontinuity.} 
These studies have demonstrated the potential of the Faraday instability to promote mixing in stably stratified fluids depending on the various parameters, such as frequency and amplitude of the vibrations and the strength of the initial stratification. Understanding the turbulent diapycnal mixing in stably stratified ocean interior is important as it plays a major role in the global overturning circulation that transports water mass, heat, nutrients, and other biochemical substances \citep{wunsch2004vertical,talley2016changes}. The Faraday instability provides a framework to understand these flows and can help to elucidate the underlying physical processes that govern mixing in the oceanographic flows. The Earth's rotation significantly affects the dynamics of diapycnal mixing processes in the thermocline, as reported in studies by  \citet{noh1990turbulent,fernando1991turbulent} and \citet{fleury1991effects}. For example, $N/f \approx 4.67$, where $N$ is the Brunt V\"{a}is\"{a}l\"{a} frequency, in the abyssal southern ocean at mid-latitude \citep{nikurashin2013routes,rosenberg2015evidence}. Modeling the properties of turbulence and the associated irreversible mixing, where external forcing, Earth's rotation and stratification play a central role, is a major challenge in geophysical fluid dynamics. Motivated by the turbulent mixing in the oceanic flows, we aim to quantify the irreversible mixing and mixing efficiencies associated with the turbulent mixing driven by the time-periodic accelerations in rotating miscible fluids using DNS. %This system provides a framework to understand the physics associated with the diapycnal mixing processes in oceanographic flows.
 In the present investigations, we follow the framework of \cite{winters1995available}, which was used to study the energetics of shear-induced mixing of stably stratified fluids by partitioning the total PE into BPE and APE. %Note that the turbulent mixing is driven by the Faraday instability in the present study, whereas it is driven by the Kelvin–Helmholtz instability in the stratified shear flows. Therefore, the dynamics of the mixing layer differ from those of stratified shear flows
 It is important to note that the Faraday instability drives the turbulent mixing, which induces small-scale turbulence and vertical motions that can enhance the mixing of stably stratified fluids. However, the turbulent mixing in stratified shear flows is typically driven by the Kelvin-Helmholtz instability resulting from velocity shear across the density gradient. As a result, the dynamics of the mixing layer differ between these two scenarios. Here, we focus on how the rotation affects the formation and subsequent breakdown of the structures on the diffuse interface to produce turbulence, irreversible mixing, and mixing efficiencies. 
%The present investigation aims to quantify the turbulent mixing driven by Faraday instability in terms of the exchanges among the total potential energy (TPE), the background potential energy (BPE), and the available potential energy (APE) under the influence of rotation.\\ 
To address these aims, the paper is organized as follows. In \cref{sec:numerical details}, we discuss the governing equations with boundary equations and the set-up for the simulations and briefly explain the theoretical framework for linear stability analysis developed in \cite{singh2022onset}. We then discuss the predictions of the linear stability analysis and the numerical results in \cref{{sec:results}}. Finally, we conclude in \cref{sec:conclusions}.

\section{Methods}\label{sec:numerical details}
\subsection{Governing equations}
The two-layer miscible fluid system with small density contrast is considered here, which is driven by vertical periodic oscillations of acceleration $g(t)=g_0(1+F\cos{(\omega t))}$. %, where $g_0$ is the mean acceleration.
We consider the density of the mixture to be a linearly varying function of mass concentration such that the lighter fluid with concentration $C(\rho_2)=0$ is placed above the denser fluid with $C(\rho_1)=1$ in a rectangular domain. The unsteady incompressible flow with Boussinesq approximation is governed by the three-dimensional conservation equations for mass, momentum, and concentration field under vertically periodic forcing $g(t)$, in the form \citep{singh2022onset} %The governing equations are the three-dimensional incompressible unsteady Navier–Stokes equations with Boussinesq approximation %and are solved in a Cartesian coordinate system on a staggered grid arrangement as discussed in \citet{singh2022onset}.
 \begin{subequations}
    \begin{equation}
    \label{continuity}
    \nabla \cdot \boldsymbol{U} = 0,
    \end{equation}
    \begin{equation}
    \label{momentum}
     \frac{\partial \boldsymbol{U}}{\partial t} + \boldsymbol{U}\cdot \nabla \boldsymbol{U} = -\nabla P - f\hat{k} \times \boldsymbol{U}  + \nu {\nabla}^2 \boldsymbol{U} - 2\mathcal{A}{C}g(t) \hat{k},
    \end{equation}
    \begin{equation}
    \label{concentration}
     \frac{\partial{C}}{\partial t} + \boldsymbol{U}\cdot \nabla {C} = \kappa {\nabla}^2 {C}.
    \end{equation}
 \end{subequations}
Here, $\boldsymbol{U}$ denotes the velocity vector with components ($U_1$, $U_2$, $U_3$) in the horizontal $x_1$, $x_2$ and vertical $x_3$ directions respectively, 
%Here, $\boldsymbol{U}(\boldsymbol{x},t)$ is the velocity field,
${P}$ is the pressure deviation from the hydrostatic background state, % $f$ is the Coriolis's frequency,
$\nu$ ($\mathrm{m}^2\, \mathrm{s}^{-1}$) is the kinematic viscosity, and $\kappa$ ($\mathrm{m}^2\, \mathrm{s}^{-1}$) is the diffusion coefficient. %, and $\mathcal{A}=\left(\rho_1-\rho_2\right)/\left(\rho_1+\rho_2\right)$ is the Atwood number expressing the density contrast between the fluids.
The Reynolds decomposition is performed to decompose a variable, say $B$, into its horizontally averaged mean component $\langle B \rangle _H$ and fluctuating component $b$, given by 

\begin{subequations}
\begin{equation}
\label{reynolds decomposition}
 B({x_1,x_2,x_3},t)=\langle B \rangle _H({x_3},t) + b({x_1,x_2,x_3},t),
 \end{equation}
 \begin{equation}
\label{average}
 \quad \langle B \rangle _H({x_3},t)= \frac{1}{l_{x_1} l_{x_2}} \int_{{-l_{x_1}}/{2}}^{{+l_{x_1}}/{2}} \int_{{-l_{x_2}}/{2}}^{{+l_{x_2}}/{2}} B ({x_1,x_2,x_3},t)\mathrm{d}x_1\mathrm{d}x_2.
\end{equation}
\end{subequations}

Here, $\langle \, \rangle _H$ denotes the horizontal average in the homogeneous directions $x_1$ and $x_2$, $\langle b \rangle _H=0$, and $l_{x_1}$, $l_{x_2}$ are the domain lengths in the $x_1$ and $x_2$ directions. The mixing zone size-$L$ is computed from mean concentration profile $\langle C \rangle_H$ \citep{andrews1990simple} as $L=6\int_{-\infty}^{+\infty} \langle C \rangle_H  (x_3,t) \left(1-\langle C \rangle_H (x_3,t)\right)\mathrm{d}x_3$. The stratification (or Brunt V\"{a}is\"{a}l\"{a}) frequency is defined as 
%$N=\left({-2\mathcal{A} g_0{\partial \langle C \rangle_H}/{\partial x_3}}\right)^{1/2}$,
 \begin{equation}
      \label{strat fre}
    N=\sqrt{-2\mathcal{A} g_0\frac{\partial \langle C \rangle_H}{\partial x_3}}, 
  \end{equation}
  where the vertical gradient of mean concentration approximated as ${\partial \langle C \rangle_H}/{\partial x_3}=-1/L$.

\subsection {Theoretical model}
\cite{singh2022onset} presents the details of the inviscid linear stability analysis of the Faraday instability in a rotating and vertically oscillating two-layer miscible fluid system. Here, we briefly discuss the origin of the derivation of the linearized governing equations (Mathieu equations) along with the assumptions. The equations for the fluctuating velocity $\boldsymbol{u}(\boldsymbol{x},t)$ and the concentration $c(\boldsymbol{x},t)$ fields are obtained by substituting the Reynolds decomposition \ref{reynolds decomposition} into the governing equations {\ref{continuity}}, {\ref{momentum}} and {\ref{concentration}}, with mean velocity field $\langle \boldsymbol{U} \rangle _H =0$ and $\langle \boldsymbol{u} \rangle _H=\langle c \rangle _H=\langle p \rangle _H=0$. The resulting equations are given as
\begin{subequations}
\begin{equation}
\label{continuity fluc}
\frac{\partial u_i}{\partial x_i}=0,
\end{equation}
\begin{equation}
\label{velocity fluc}
\frac{\partial u_i}{\partial t} + u_j \frac{\partial u_i}{\partial x_j} =  - \frac{\partial p}{\partial x_i}  + f\epsilon_{ij3} u_j\hat{e}_3 - 2\mathcal{A}g(t)c\delta_{i3} + \frac{\partial \langle u_i u_3 \rangle_H}{\partial x_3}+ \nu {\nabla}^2 u_i, 
\end{equation}
\begin{equation}
\label{concentration fluc}
\frac{\partial c}{\partial t} + u_j \frac{\partial c}{\partial x_j} = - u_3 \frac{\partial \langle C \rangle_H}{\partial x_3} +  \frac{\partial \langle u_3 c \rangle_H}{\partial x_3} + \kappa\, {\nabla}^2 c. 
\end{equation}
\end{subequations}

We assume that the fluids are inviscid and the fluctuations in velocity and concentration are small. Therefore, the non-linear, the viscous, and the diffusion terms are neglected from the equations \ref{velocity fluc} and \ref{concentration fluc}. The concentration profile is assumed to be piecewise, i.e. constant in the bottom and upper pure (unmixed) fluids and linear in the mixing layer, with a vertical gradient of mean concentration ${\partial \langle C \rangle_H}/{\partial x_3}=0$, in the unmixed region and ${\partial \langle C \rangle_H}/{\partial x_3}=\text{constant}$, in the mixed region. The detailed derivation of the linearized equations and their solution strategy is presented in \cite{singh2022onset}. The final system of linearized equations (for $N>f$), equivalent to a set of Mathieu equations, is given as
 \begin{equation}
     \label{a0 eq3}
     \frac{\partial^2 a}{\partial \tau^2} + \left( \frac{f^2 \cos^2{(\theta)}}{\omega^2} +  \frac{N^2 \sin^2{(\theta)}}{\omega^2} \left(1+F\cos{(\tau)} \right) \right) a = 0,
 \end{equation} 
  where, $a$ represents the amplitude of the concentration field mode, $\tau\,(=\omega t)$ is the non-dimensional time, and $\theta$ is the angle between the vertical axis and the wavevector $\mathbf{K}$ with components $(k,l,m)$ in the $x_1$, $x_2$ and $x_3$ directions, respectively. The Floquet theory, which provides the solutions to linear differential equations with periodic coefficients, is utilized to solve a set of Mathieu equations \ref{a0 eq3}. Details of solving the equation \ref{a0 eq3} using Floquet theory are presented in Appendix \ref{appA}. %We solve the system of  equations using the procedure outlined in the appendix of
  The solutions of the equation \ref{a0 eq3} can be stable or unstable which governs the stability of the system. 

 \begin{table}
  \begin{center}
\def~{\hphantom{0}}
\setlength{\tabcolsep}{8pt} % Default value: 6pt
\renewcommand{\arraystretch}{1.2} % Default value: 1
  \begin{tabular}{lcccccccc}
       Case        & $F$      & $\omega \:(\mathrm{rad\:s^{-1}})$    & $f\:(\mathrm{s^{-1}})$   & $f/\omega$ & $\left(f/\omega \right)^2$ & $\eta_{cu}$  \\[3pt]
       F075f/$\omega0$  & 0.75 & 0.67   & 0     & 0  &   0   & $\sim$ 0.495\\
       F075f/$\omega$48      & 0.75 & 0.67   & 0.322 & 0.481   & 0.23   & $\sim$ 0.363\\
       F075f/$\omega$59      & 0.75 & 0.67   & 0.396 & 0.591   & 0.35   & never saturate\\
       F1f/$\omega0$  & 1.0 & 0.7   & 0     & 0  &   0   & $\sim$ 0.421\\
       F1f/$\omega$48      & 1.0 & 0.7   & 0.338 & 0.482   & 0.23   & $\sim$ 0.372\\
       F1f/$\omega$59      & 1.0 & 0.7   & 0.414 & 0.591   & 0.35   & never saturate\\
       F2f/$\omega0$  & 2.0 & 0.8   & 0     & 0  &   0   & $\sim$ 0.392\\
       F2f/$\omega$48      & 2.0 & 0.8   & 0.384 & 0.48   & 0.23   & $\sim$ 0.366\\
       F2f/$\omega$59      & 2.0 & 0.8   & 0.473 & 0.592   & 0.35   & never saturate\\
       F3f/$\omega0$  & 3.0 & 0.9   &   0   & 0  &   0   & $\sim$ 0.345\\
       F3f/$\omega$48      & 3.0 & 0.9   & 0.432 & 0.48   & 0.23   & $\sim$ 0.333\\
       F3f/$\omega$59      & 3.0 & 0.9   & 0.532 & 0.592   & 0.35   & never saturate\\
      
  \end{tabular}
  \caption{Simulation parameters: $\eta_{cu}$ is the  cumulative mixing efficiency. For all cases we use Atwood number $\mathcal{A}=0.01$, %, initial mixing zone width $L_0=0.096 \:\mathrm{m}$,
  kinematic viscosity $\nu=1\times10^{-4}\:\mathrm{m^2\:s^{-1}} $, diffusion coefficient $\kappa=1\times10^{-4}\:\mathrm{m^2\:s^{-1}}$, and $g_0=10\:\mathrm{m\:s^{-2}}$ \citep{briard2019harmonic,singh2022onset}.% Domain size: $l_{x_1}=l_{x_2}=2\pi \:\mathrm{m}$, and $l_{x_3}=2H=3.5\pi \:\mathrm{m}$. %including the sponge region of thickness $H_s=0.64\pi \:\mathrm{m}$ at top and bottom boundaries. 
  %Grid points: $N_{x_1}=N_{x_2}=512$ (uniform), and $N_{x_3}=512$ (non-uniform with clustering at the centre region of thickness $1.53\pi \:\mathrm{m}$ with $\Delta x_{3_{min}}=\Delta x_1=\Delta x_2$).
  }  \label{tab:parameters}
  \end{center}
\end{table}

\subsection {Simulation details}
 The governing equations {\ref{continuity}}, {\ref{momentum}}, and {\ref{concentration}} are solved in a Cartesian coordinate system on a staggered grid using a finite difference method. The numerical algorithms are presented in \citet{singh2022onset}. The numerical solver has been extensively validated and used for studies of rapidly rotating convection-driven dynamos \citep{naskar2022direct, naskar_pal_2022}, rotating convection \citep{pal2020evolution}, and several stratified free-shear and wall-bounded turbulent flows \citep{brucker2010comparative, pal2013spatial, pal2015effect, pal2020deep}. A sponge region is used near the top and bottom boundaries to control the spurious reflections from the disturbances propagating out of the domain, where damping functions gradually relax the velocities to their corresponding values at the boundaries. These damping functions are added on the right-hand side of the momentum equation \ref{momentum} as explained in \citet{brucker2010comparative}. The sponge region is sufficiently far away from the mixing region such that it does not affect the dynamics of the mixing of fluids. We consider periodic boundary conditions for all variables in the $x_1$ and $x_2$ (horizontal) directions, whereas at the top and bottom walls ($x_3$), no-slip boundary conditions for velocity vector $\boldsymbol{U}$ and Neumann boundary conditions for $C$ and pressure $P$ are used. The initial concentration profile \citep{briard2019harmonic,briard2020turbulent,singh2022onset} in the present simulations is given as
 \begin{equation}
 \label{conc_profile}
     C\left(\boldsymbol{x},t=0 \right)=\frac{1}{2}\left(1-\tanh{\left(\frac{2x_3}{\delta}\right)}\right),
 \end{equation}
where the parameter $\delta$ is used to change the initial mixing zone size-$L_0$. $\delta=0.035$ and corresponding $L_0=0.096 \:\mathrm{m}$ is used across all simulations. Broadband fluctuations are imposed on the initial concentration profile \ref{conc_profile} as reported in \citet{singh2022onset}. The domain size in the horizontal directions is $l_{x_1}=l_{x_2}=2\pi \:\mathrm{m}$, and in the vertical direction is $l_{x_3}=2H=3.5\pi \:\mathrm{m}$ which includes sponge region of thickness $H_s=0.64\pi \:\mathrm{m}$ at top and bottom boundaries. We use uniform grids $N_{x_1}=N_{x_2}=512$ in the horizontal directions, while in the vertical direction $x_3$, non-uniform grids $N_{x_3}=512$ are used with clustering at the vertical centre region of thickness $1.53\pi \:\mathrm{m}$ with $\Delta x_{3_{min}}=\Delta x_1=\Delta x_2$. Table \ref{tab:parameters} shows the key parameters for all simulation cases. We select three cases at each $F$ depending on the value of $f/\omega$, where sub-harmonic instability saturates for $f/\omega=0$, and 0.48 ($\left(f/\omega \right)^2<0.25$) and never saturates for $f/\omega=0.59$ ($\left(f/\omega \right)^2\geq0.25$) \citep{singh2022onset}. We refer each case with a unique name for example F075f/$\omega$48, which indicates that $F=0.75$ and $f/\omega=0.48$.

\section{Results}\label{sec:results}
\subsection{Linear stability analysis }\label{sec:linearresults} 
%  \captionsetup[subfigure]{textfont=normalfont,singlelinecheck=off,justification=raggedright}
%  \begin{figure}
%  \centering
%  %\begin{subfigure}{0.485\textwidth}
%  \begin{subfigure}{0.425\textwidth}
% 	\centering
% 	\includegraphics[width=1.0\textwidth,trim={0cm 0.2cm 0cm 0cm},clip]{figures/3d_F1_p2.pdf}
% 	\caption{}	\label{subfig:3dF075}
%  \end{subfigure}
%  \quad
%  %	\hfill
% % 	\begin{subfigure}{0.485\textwidth}
% % 		\centering
% % 		\includegraphics[width=1.0\textwidth,trim={0cm 0.2cm 0cm 0cm},clip]{figures/3d_F1_p2.pdf}
% % 		\caption{}	\label{subfig:3dF1}% 	
% % 	\end{subfigure}
% % 	\begin{subfigure}{0.485\textwidth}
% % 		\centering
% % 		\includegraphics[width=1.0\textwidth,trim={0cm 0.2cm 0cm 0cm},clip]{figures/3d_F2_p2.pdf}
% % 		\caption{}	\label{subfig:3dF2}% 	
% % 	\end{subfigure}
% % \hfill
%  \begin{subfigure}{0.425\textwidth}
% 	\centering
% 	\includegraphics[width=1.0\textwidth,trim={0cm 0.2cm 0cm 0cm},clip]{figures/3d_F3_p2.pdf}
% 	\caption{}	\label{subfig:3dF3}
%  \end{subfigure}
%  \caption{Top views of the three-dimensional Mathieu stability diagram \citep{singh2022onset} at forcing amplitudes (\textit{a}) $F=1$ and (\textit{b}) $F=3$. The stable (white) regions are in between the unstable red (sub-harmonic) and cyan (harmonic) colored tongues. The horizontal green line segment denotes the case without rotation ($f=0$) whereas rotation cases ($f\neq0$) are indicated by the inclined blue, pink, black, purple and orange line segments.}	\label{fig:3dFplane}
% \end{figure}

  \captionsetup[subfigure]{textfont=normalfont,singlelinecheck=off,justification=raggedright}
 \begin{figure}
 \centering
 \begin{subfigure}{0.495\textwidth}
	\centering
	\includegraphics[width=1.0\textwidth,trim={0cm 0.05cm 0.2cm 0cm},clip]{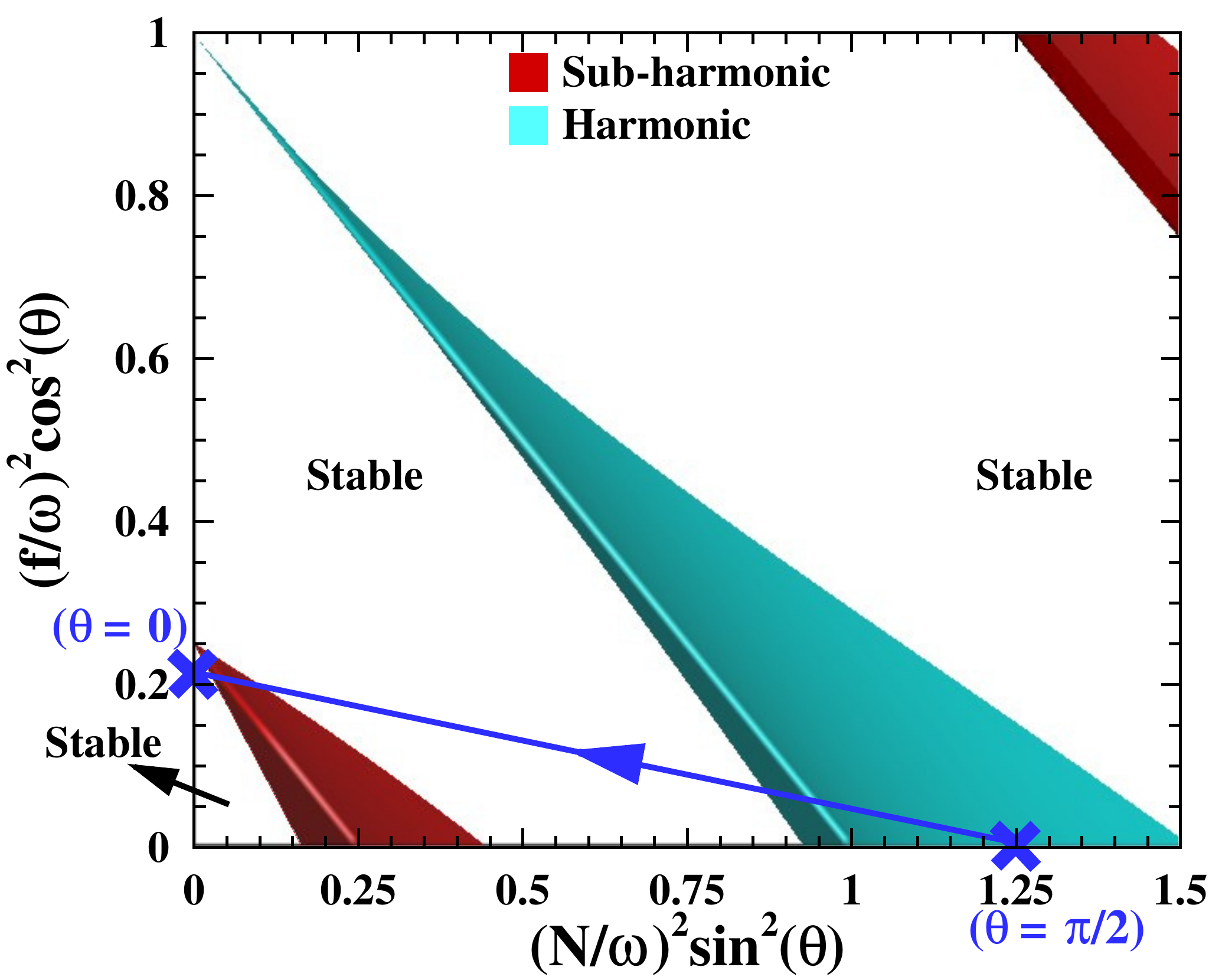}
	\caption{}	\label{subfig:3dF1a}
 \end{subfigure}
 %\quad
  \hfill
  \begin{subfigure}{0.495\textwidth}
	\centering
	\includegraphics[width=1.0\textwidth,trim={0cm 0.05cm 0.2cm 0cm},clip]{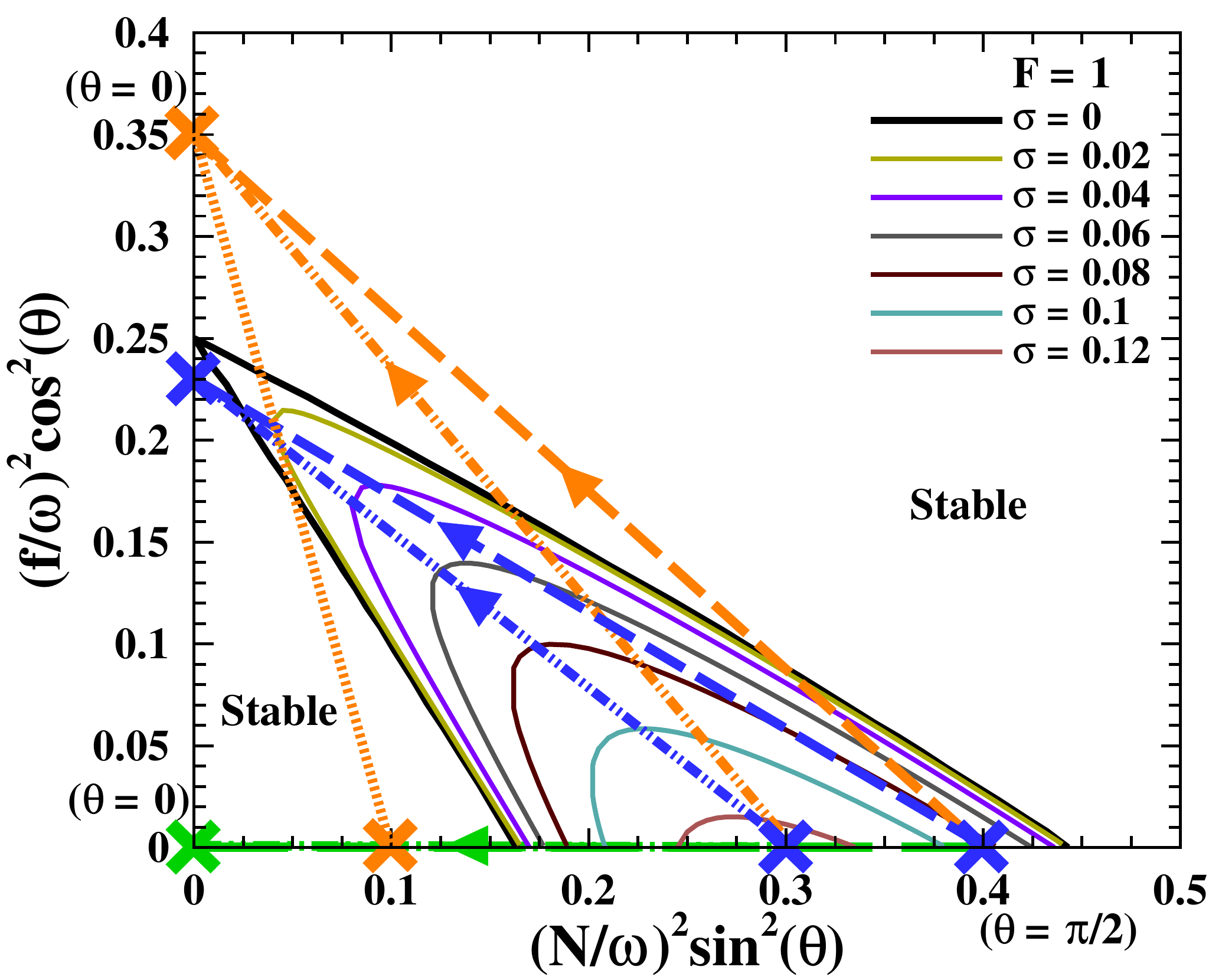}
	\caption{}	\label{subfig:3dF1b}
 \end{subfigure}
 \begin{subfigure}{0.495\textwidth}
	\centering
	\includegraphics[width=1.0\textwidth,trim={0cm 0.05cm 0.05cm 0cm},clip]{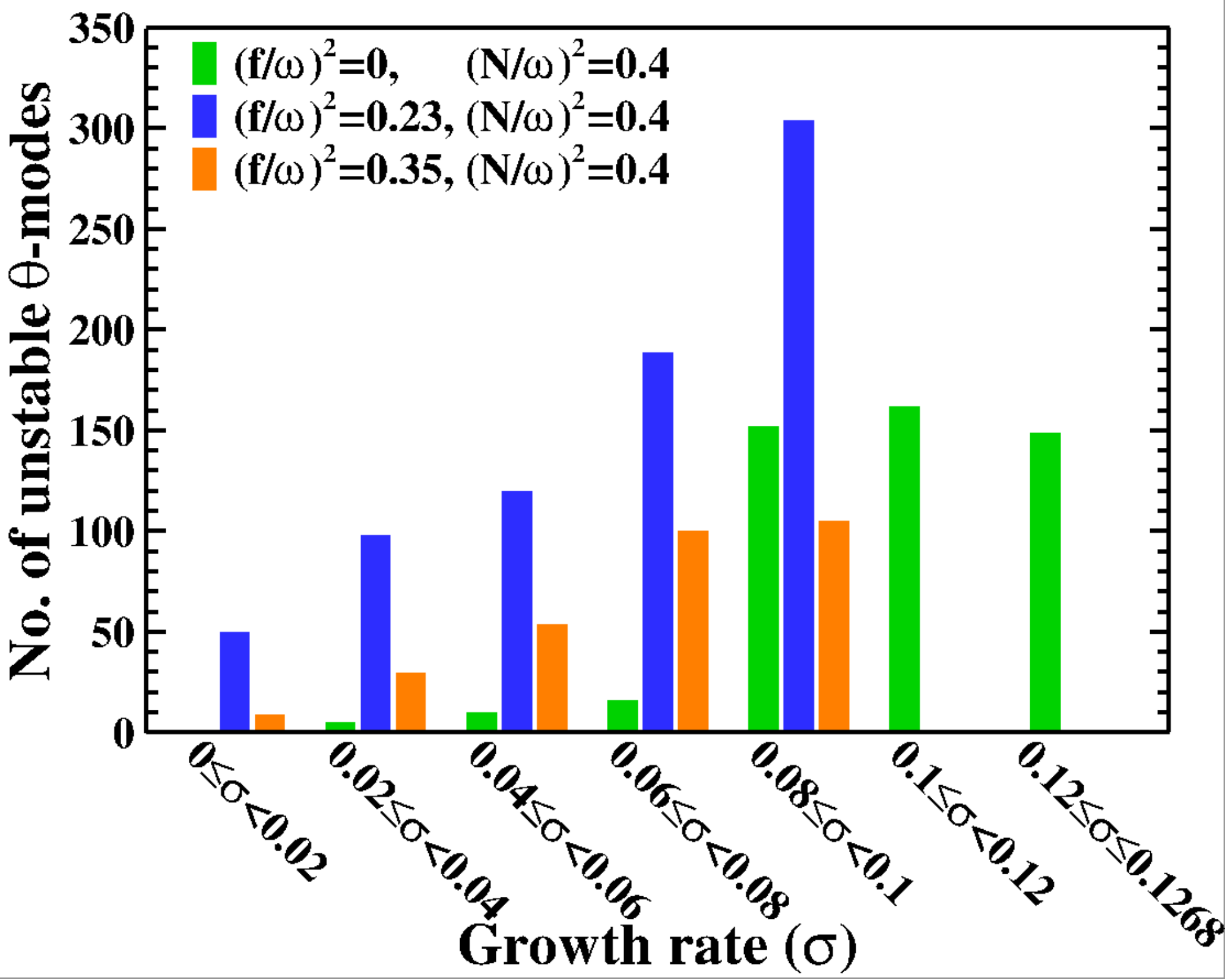}
	\caption{}	\label{subfig:F1_N2W2_04}
 \end{subfigure}
 \hfill
  \begin{subfigure}{0.495\textwidth}
	\centering
	\includegraphics[width=1.0\textwidth,trim={0cm 0.05cm 0.05cm 0cm},clip]{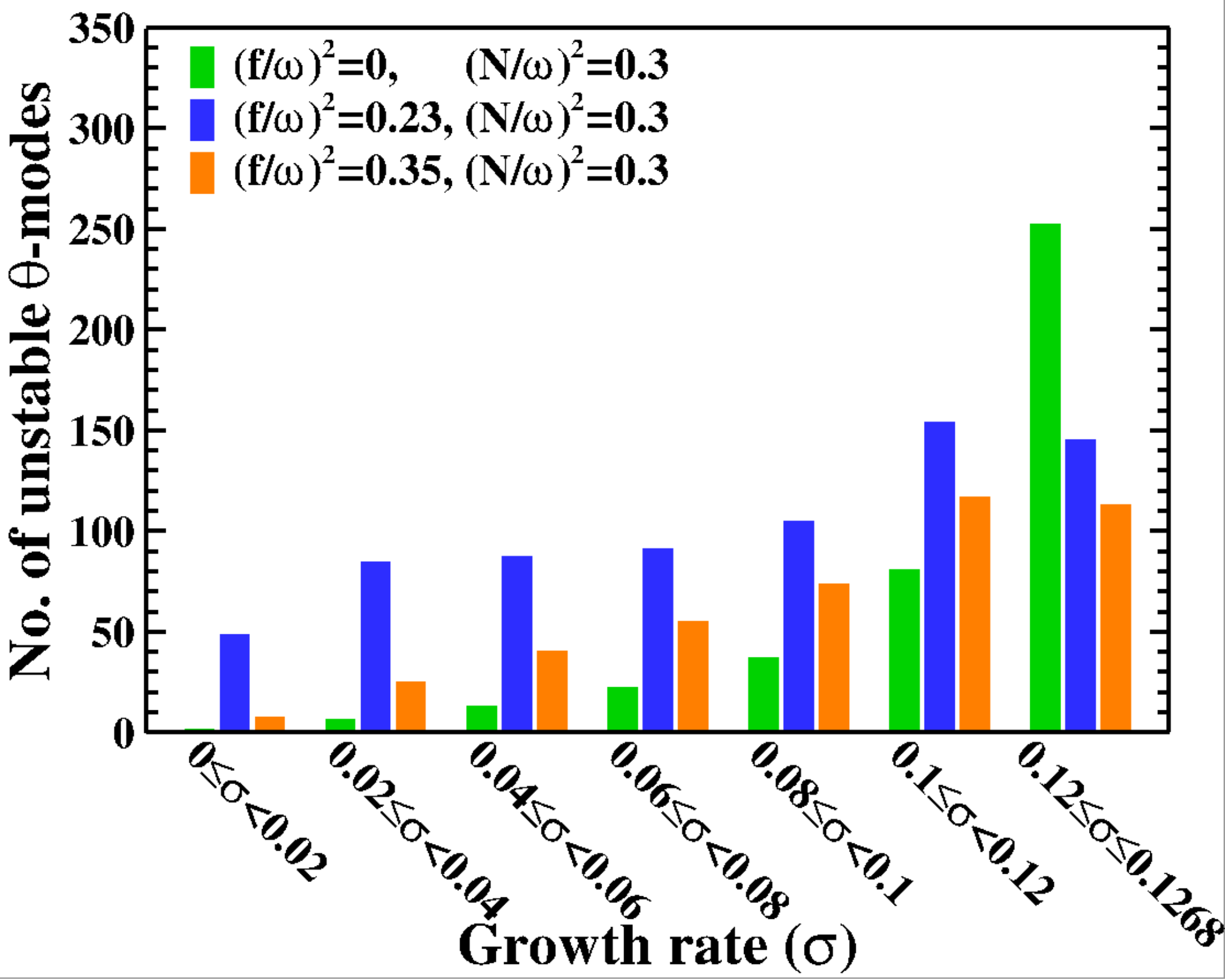}
	\caption{}	\label{subfig:F1_N2W2_03}
 \end{subfigure}

%  \begin{subfigure}{0.495\textwidth}
% 	\centering
% 	\includegraphics[width=1.0\textwidth,trim={0cm 0.05cm 0.2cm 0cm},clip]{figures/F1_N2W2_02.pdf}
% 	\caption{}	\label{subfig:F1_N2W2_02}
%  \end{subfigure}
%  \begin{subfigure}{0.495\textwidth}
% 	\centering
% 	\includegraphics[width=1.0\textwidth,trim={0cm 2cm 0.2cm 0cm},clip]{figures/F1_N2W2_num.pdf}
% 	\caption{}	\label{subfig:F1 N2W2 num}
%  \end{subfigure}
%   \hfill
%  \begin{subfigure}{0.495\textwidth}
% 	\centering
% 	\includegraphics[width=1.0\textwidth,trim={0cm 2cm 0.2cm 0cm},clip]{figures/F1_TKE.pdf}
% 	\caption{}	\label{subfig:F1 tke}
%  \end{subfigure}
 \caption{(\textit{a}) Stability diagram for the solutions to the Mathieu equation \ref{a0 eq3} \citep{singh2022onset} at forcing amplitude $F=1$. The stable (white) regions are in between the unstable red (sub-harmonic) and cyan (harmonic) colored tongues. The inclined blue line (solid) segment represents the frequencies excited, corresponding to an angle $\theta \in [0,\pi/2]$, for a given $\left(f/\omega \right)^2=0.23$ (left end ($\times$) at $\theta=0$), and initial condition $\left(N_0/\omega \right)^2$ (right end ($\times$) at $\theta=\pi/2$) in the first harmonic tongue. Arrow indicates the evolution of $\left(N/\omega \right)^2$ as the mixing zone size-$L$ increases. Panel (\textit{b}) is an enlargement of the first leftmost sub-harmonic tongue in (\textit{a}). The black curve denotes the neutral stability curve for the growth rate $\sigma=0$. The color curves respectively denote the contours of different growth rates ($\sigma$) at intervals of $0.02$. The dashed and dash-dotted green, blue and orange segments demonstrate the non-rotating case $\left(f/\omega \right)^2=0$ and rotating cases $\left(f/\omega \right)^2=0.23$ and 0.35, respectively, when $\left(N/\omega \right)^2$ reaches to 0.4 and 0.3 during the growth of $L$. % For a given $\left(f/\omega \right)^2$ and $\left(N/\omega \right)^2$, these segments (left end ($\times$); $\theta=0$ and right end ($\times$); $\theta=\pi/2$) represent the frequencies $\in[\left(f/\omega \right)^2, \left(N/\omega \right)^2]$ excited, corresponding to an angle $\theta \in [0,\pi/2]$. 
 The possible number of unstable $\theta$-modes permitted to grow at various growth rates ($\sigma$) for each $\left(f/\omega\right)^2=$ 0, 0.23, and 0.35 cases are shown in (\textit{c}) when $\left(N/\omega \right)^2=0.4$, and (\textit{d}) when $\left(N/\omega \right)^2=0.3$. }
 \label{fig:growth rates}
\end{figure}

Figure \ref{subfig:3dF1a} shows the stability diagram at forcing amplitude $F=1$ in the %$\left(\left(N/\omega \right)^2\sin^2(\theta)-\left(f/\omega \right)^2 \cos^2(\theta) \right)$ 
$\left(N/\omega \right)^2-\left(f/\omega \right)^2$ parameter plane, obtained by solving the linearized governing equations (Mathieu equations) \ref{a0 eq3}. The stable (white) regions are in between the unstable red (sub-harmonic) and cyan (harmonic) colored tongues. %In the sub-harmonic tongue, the solutions grow and oscillate with possible frequencies of an odd multiple of the $\omega/2$. The solutions inside the harmonic tongue grow with frequencies that are integer multiples of $\omega$
%{\color{red} Probably include this in the main text} %{\color{blue}``$cos(\tau)$ in equation 2.5 is $2\pi$ ($T=2\pi$) periodic i.e. $\omega = 2\pi f = 2 \pi / T = 2 \pi / 2 \pi = 1$. From the Floquet analysis, solutions in the first sub-harmonic tongue are $4\pi$ periodic (fre = 1/2 or $\omega/2$), and in the other sub-harmonic tongues they are like $3\omega/2$, $5\omega$/2). Therefore the solutions grow and oscillate with possible frequencies of an odd multiple of the $\omega/2$ in the sub-harmonic tongues. On the other hand, solutions in the first harmonic tongue are $2\pi$ (fre = 1 or $\omega$), and in the other harmonic tongues, they are $2\omega, 3\omega$ ...). We can only mention the solutions in the first sub-harmonic and harmonic tongue.''}.
From the Floquet analysis (see Appendix \ref{appA}, \cite{jordan2007nonlinear} and \cite{singh2022onset}), the solutions inside the first sub-harmonic tongue are periodic with period $T=4\pi$. Therefore, the solutions grow and oscillate with frequency $\omega/2$ because $\mathrm{cos}(\tau)$, where $\tau=\omega t$, in the equation (\ref{a0 eq3}) is periodic with $T=2\pi$ and frequency is  $2\pi/T=1$. However, the solutions grow in the second and third sub-harmonic tongues with frequencies of $3\omega/2$ and $5\omega$/2, respectively. The solutions inside the first harmonic tongue are $2\pi$ periodic and grow with frequency $\omega$, whereas, in the second and third harmonic tongues, the solutions grow with frequencies of $2\omega$ and $3\omega$, respectively. The solutions that lie outside the harmonic and sub-harmonic instability tongues are stable. The contour curves for the growth rates ($\sigma$), range from $\sigma=0.02$ to $\sigma=0.12$ with a step size of 0.02, of the unstable $\theta$-modes inside the first leftmost sub-harmonic tongue are illustrated in figure \ref{subfig:3dF1b}. The black curve corresponds to the neutral stability curve for $\sigma=0$, which separates the regions of stable and unstable solutions. %The region bounded by the black curve represents the unstable tongue where the solutions grow and oscillate with possible frequencies of odd multiple of the $\omega/2$ which is half of the external forcing frequency. Therefore this unstable tongue is referred as sub-harmonic instability tongue and correspond to the first left most sub-harmonic (red) tongue shown in the inset of figure \ref{subfig:3dF1}. The contour curves for various growth rates ($\sigma$), from $\sigma=0.02$ to $\sigma=0.12$ with a step size of 0.02, are shown inside the sub-harmonic tongue with different colors. Other unstable harmonic tongue (cyan) can be seen in the inset of figure \ref{subfig:3dF1} and the solutions inside this tongue grow with frequencies that are integer multiples of $\omega$. The solutions that lie outside the harmonic and sub-harmonic instability tongues are stable. 
The inclined blue line (solid) segment in figure \ref{subfig:3dF1a} demonstrates the possible frequencies ($\in[\left(f/\omega \right)^2, \left(N_0/\omega \right)^2]$) that can be excited corresponding to an angle $\theta \in [0,\pi/2]$, for a given $\left(f/\omega \right)^2=0.23$ and initial stratification condition $\left(N_0/\omega \right)^2$ in the first harmonic tongue. This segment ranges from $\left(f/\omega \right)^2$ (left end ($\times$) at $\theta=0$) to $\left(N_0/\omega \right)^2$ (right end ($\times$) at $\theta=\pi/2$). The unstable $\theta$-modes inside the sub-harmonic or harmonic tongues trigger the Faraday instability. This results in the growth of mixing zone size-$L$ and decrease of $\left(N/\omega \right)^2$, as indicated by the arrow directing to the left since $\left(N/\omega \right)^2=2\mathcal{A}g_0/\left(L \omega^2\right) \propto 1/L$ from equation \ref{strat fre}. Therefore, the condition for the occurrence of instabilities \citep{grea2018final,singh2022onset} is defined based on the fact that if at least one $\theta$-mode falls in any unstable tongues solution will grow, resulting in instability. This condition is given as
     \begin{equation}
      \label{inst condition2}
        \left(\frac{N}{\omega} \right)^2>\mathcal{G}(f,F) \;,
     \end{equation}
 where $\mathcal{G}(f, F)$ denotes the extreme left boundary of the first leftmost sub-harmonic tongue (see figure \ref{subfig:3dF1a}). With time evolution, $L$ grows due to unstable $\theta$-modes and the right end ($\left(N/\omega \right)^2$) of the blue line (solid) segment (figure \ref{subfig:3dF1a}) enters the first sub-harmonic tongue as indicated by the blue $\times$ at $\left(N/\omega \right)^2=0.4$ in figure \ref{subfig:3dF1b}. At this instant, many frequencies ($\in[\left(f/\omega \right)^2, \left(N/\omega \right)^2]$) are excited, corresponding to an angle $\theta \in [0,\pi/2]$ and growth rates $\sigma$, as depicted by the inclined dashed blue segment in figure \ref{subfig:3dF1b}. The sub-harmonic instabilities are triggered in this regime resulting in the intense growth of $L$ \citep{singh2022onset}. The process repeats, and $\left(N/\omega \right)^2$ moves to 0.3, as indicated by the dash-dotted blue segment. The instability finally saturates when $\left({N}/{\omega} \right)^2$ crosses the extreme left neutral stability curve (black) i.e., $\left({N}/{\omega}\right)^2=\mathcal{G}(f,F)$. The instability condition \ref{inst condition2} is no longer fulfilled, and $L$ saturates. The horizontal green and inclined orange segments (both dash and dash-dotted) represent the cases without rotation $\left(f/\omega \right)^2=0$ and with rotation $\left(f/\omega \right)^2=0.35$, respectively. We note here that the instability saturates for $\left(f/\omega\right)^2<0.25$. An interesting feature for $\left(f/\omega\right)^2\geq0.25$ is observed when $\left(N/\omega \right)^2$ moves into the leftmost stable regime as indicated by the dotted orange segment (inclined). This segment must always pass through the first sub-harmonic tongue. The instability condition \ref{inst condition2} is always satisfied, and $L$ will continue to grow. Therefore, the instability will never saturate.\\%, resulting in the sustenance of turbulence, as shown later in figure \ref{subfig:tkeF1}. \\

 %Figure \ref{subfig:F1_N2W2_04} shows the number of unstable $\theta$-modes that fall within an interval of growth rates ($\sigma$) of 0.02-size, when $\left(N/\omega \right)^2=0.4$ for each $\left(f/\omega\right)^2=$ 0, 0.23, and 0.35 cases. 
 Figure \ref{subfig:F1_N2W2_04} depicts the possible number of unstable $\theta$-modes permitted to grow at various growth rates $\sigma$, when $\left(N/\omega \right)^2=0.4$ for each $\left(f/\omega\right)^2=$ 0, 0.23, and 0.35 cases. Here, the green, blue, and orange bars correspond to the dashed green, blue, and orange segments are shown in figure \ref{subfig:3dF1b}. For a given $\left(f/\omega\right)^2$ and $\left(N/\omega\right)^2$, we assume a step size of $0.1$ between the $\theta$ angles while calculating the number of $\theta$-modes between different growth rates with $0.02$ size intervals, except for the last interval, which spans from 0.12 to 0.1268 and contains the maximum growth rate. In the non-rotating case $\left(f/\omega \right)^2=0$, the growth rate of the fastest growing $\theta$-modes is $0.12\leq\sigma\leq0.1268$ (see figure \ref{subfig:F1_N2W2_04}). However, the growth rate of the fastest growing modes in the rotating case $\left(f/\omega \right)^2=0.23$ ($0.08\leq\sigma<0.1$) is less compared to the non-rotating case. Therefore, the mixing zone size-$L$ will increase (or $\left(N/\omega \right)^2$ decrease) gradually for $\left(f/\omega \right)^2=0.23$, than the rapid increase of $L$ (or decrease of $\left(N/\omega \right)^2$) for the non-rotating case. This results in a longer sub-harmonic instability phase for rotating case $\left(f/\omega \right)^2=0.23$ as compared to the non-rotating case. The sub-harmonic instability phase is the time interval between the onset and the saturation of the sub-harmonic instability. At a higher rotation rate $\left(f/\omega \right)^2=0.35$, fewer fastest growing $\theta$-modes have a growth rate $0.8\leq\sigma<0.1$ as compared to $\left(f/\omega \right)^2=0.23$ (see a comparison among blue and orange bars in figure \ref{subfig:F1_N2W2_04}). This demonstrates that $L$ increases much more gradually for $\left(f/\omega \right)^2=0.35$ than $\left(f/\omega \right)^2=0.23$. Figure \ref{subfig:F1_N2W2_03} illustrates that at a later instant, when $\left(N/\omega \right)^2=0.3$ and $\left(f/\omega\right)^2=0$, only a small number of $\theta$-modes are permitted to grow at smaller growth rates $\sigma<0.08$, whereas a large number of unstable modes grows at larger growth rates $\sigma\geq0.08$. The scenario is different for $\left(f/\omega\right)^2=0.23$, where a significant number of unstable modes are permitted to grow at all growth rates. %(see a comparison among green and blue bars in figure \ref{subfig:F1_N2W2_03}).
 This results in the triggering of more sub-harmonic instabilities and suggests an increase in turbulent kinetic energy ($t.k.e.$) for cases with $0<\left(f/\omega\right)^2<0.25$ as compared to the non-rotating case since sub-harmonic instabilities are responsible for the turbulent mixing \citep{singh2022onset}. Although, more unstable modes are permitted to grow at growth rates upto $\sigma<0.12$ for $\left(f/\omega\right)^2=0.35$ compared to $\left(f/\omega\right)^2=0$, a significantly less number of fastest growing modes grow at $0.12\leq\sigma\leq0.1268$. This suggests that higher rotation rates, $\left(f/\omega\right)^2\geq0.25$, have weaker turbulence. \\
 
\begin{figure}
 \centering
 
   \begin{subfigure}{0.495\textwidth}
	\centering
	\includegraphics[width=1.0\textwidth,trim={0cm 0.05cm 0.2cm 0cm},clip]{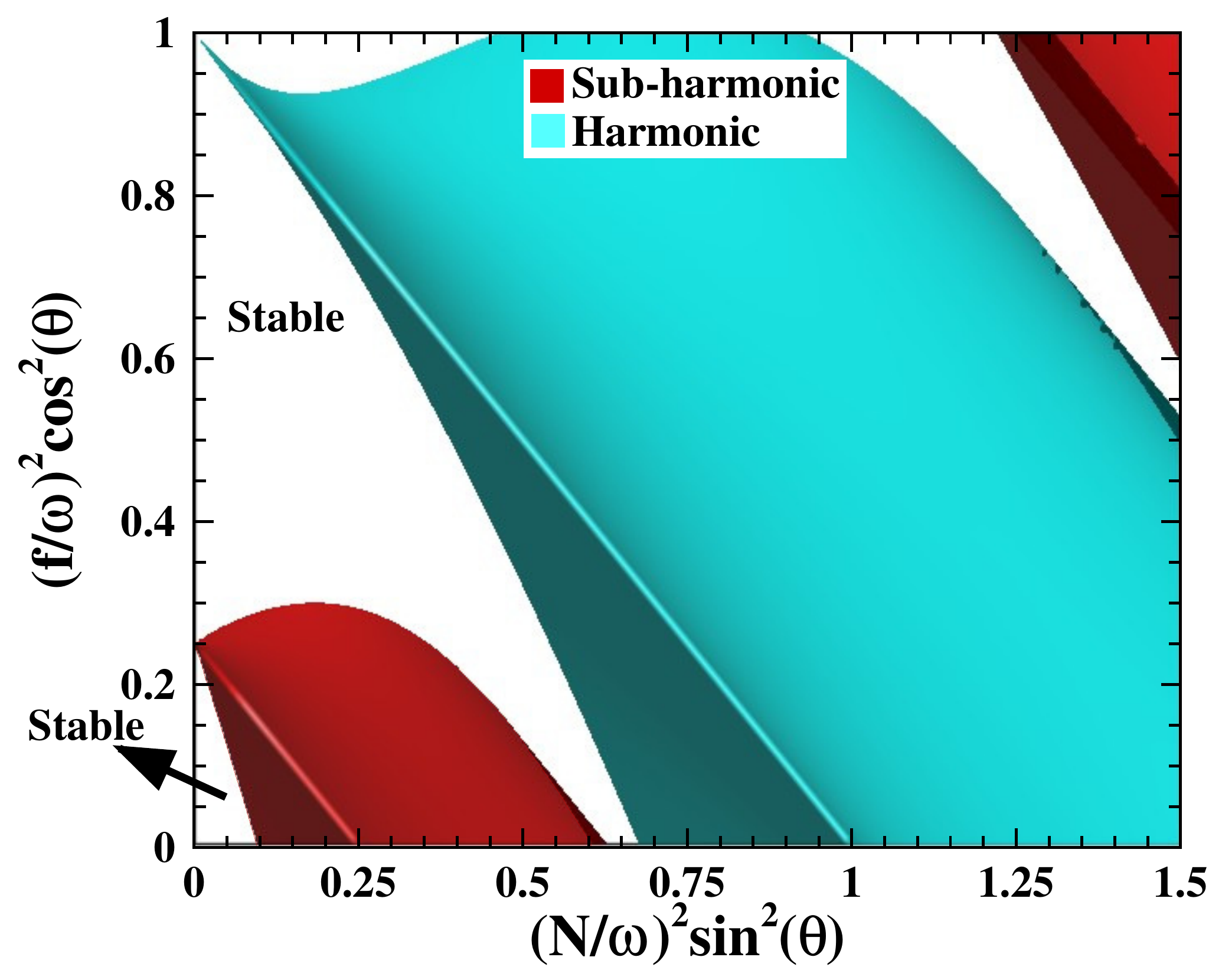}
	\caption{}	\label{subfig:3dF3a}
 \end{subfigure}
 \hfill
 \begin{subfigure}{0.495\textwidth}
	\centering
	\includegraphics[width=1.0\textwidth,trim={0cm 0.05cm 0.2cm 0cm},clip]{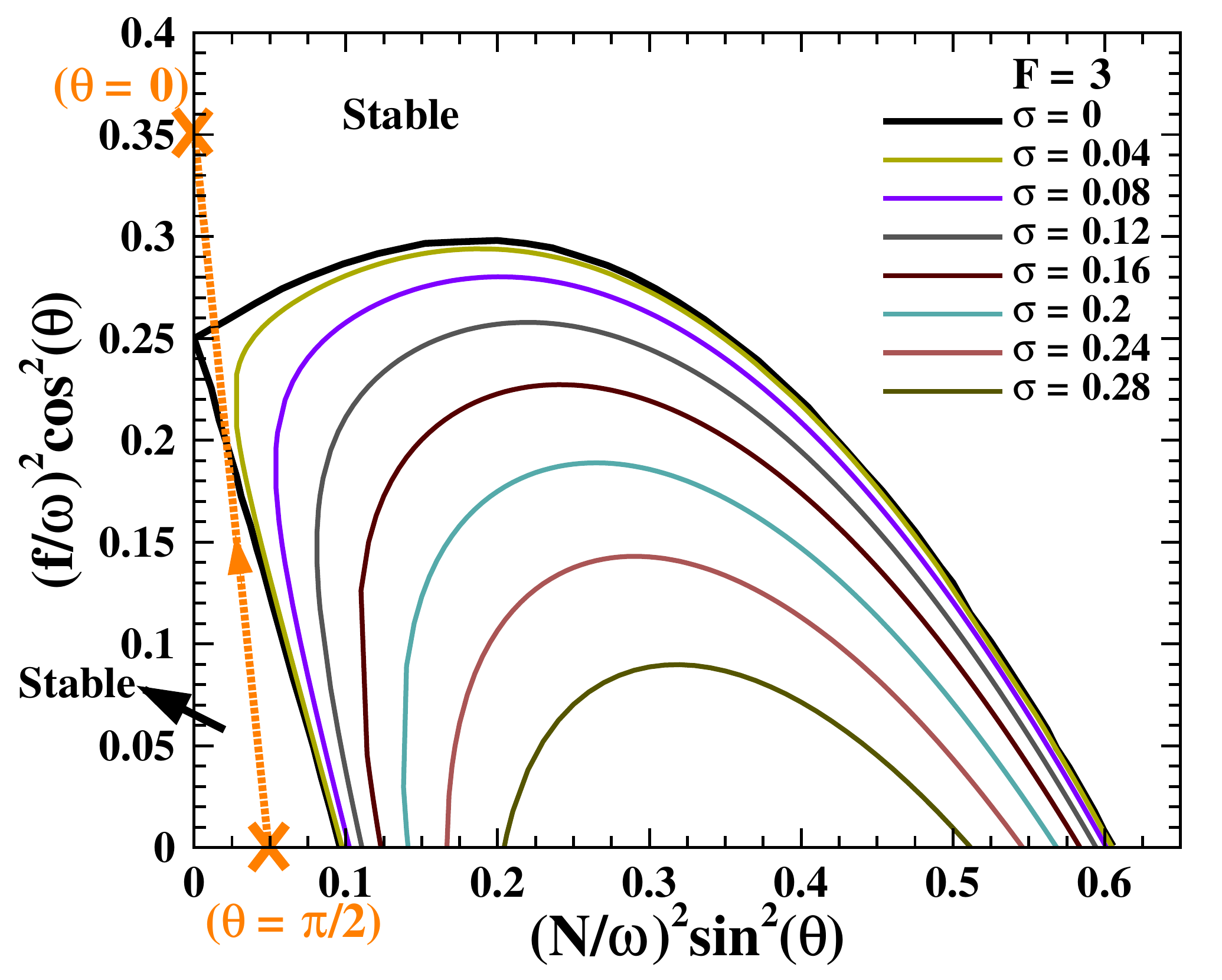}
	\caption{}	\label{subfig:3dF3b}
 \end{subfigure}
 %\hfill
%   \begin{subfigure}{0.495\textwidth}
% 	\centering
% 	\includegraphics[width=1.0\textwidth,trim={0cm 0.05cm 0.2cm 0cm},clip]{figures/F3_N2W2_03.pdf}
% 	\caption{}	\label{subfig:F3_N2W2_03}
%  \end{subfigure}
%   \hfill
%  \begin{subfigure}{0.495\textwidth}
% 	\centering
% 	\includegraphics[width=1.0\textwidth,trim={0cm 0.05cm 0.2cm 0cm},clip]{figures/F3_TKE.pdf}
% 	\caption{}	\label{subfig:F3 tke}
%  \end{subfigure}
 \caption{(\textit{a}) Stability diagram for the solutions to the Mathieu equation \ref{a0 eq3} \citep{singh2022onset} at $F=3$. Panel (\textit{b}) is the enlargement of the first leftmost unstable sub-harmonic (red) tongue of panel (\textit{a}). The unstable sub-harmonic region is bounded by the black neutral stability curve for the growth rate $\sigma=0$ and the color curves represent the contours of different growth rates ($\sigma$) at intervals of $0.04$. The dotted orange segment demonstrate the rotating case $\left(f/\omega \right)^2=0.35$. The arrow indicates the evolution of $\left(N/\omega \right)^2$ as the mixing zone size-$L$ increases.}
 \label{fig:growth rates F3}
\end{figure}

 Figure \ref{subfig:3dF3a} illustrates the stability diagram at forcing amplitude $F=3$. The contour curves for the growth rates, from $\sigma=0$ to $\sigma=0.28$ with a step size of 0.04, in the first leftmost sub-harmonic tongue, are shown in figure \ref{subfig:3dF3b}. We can observe that with an increase in $F$, the first sub-harmonic and harmonic tongues become wider, and the stable region between these tongues shrinks resulting in the early onset of the sub-harmonic instability and turbulent mixing similar to the non-rotating case \citep{singh2022onset}. Additionally, for all $\left(f/\omega\right)^2$ cases, the maximum growth rate ($\sigma_{max}$) of the fastest growing modes is $2.7$ times greater at $F=3$ than at lower forcing amplitude ($F=1$) i.e., $\sigma_{max}=0.338$ at $F=3$, whereas $\sigma_{max}=0.1268$ at $F=1$. As a result, the sub-harmonic instability is triggered earlier at $F=3$ than for $F=1$. This suggests that the sub-harmonic instability phase at $F=3$ is shorter than the sub-harmonic instability phase at $F=1$ for $\left(f/\omega\right)^2<0.25$ cases. Similar to the cases at $F=1$, the instability saturates for $\left(f/\omega\right)^2<0.25$, whereas the instability never saturates for $\left(f/\omega\right)^2\geq0.25$ and $L$ continues to grow at $F=3$, owing to the continuous triggering of the unstable $\theta$-modes as demonstrated by the dotted orange segment in figure \ref{subfig:3dF3b}. These predictions from the linear stability analysis will further support our numerical simulations as presented in the next section.\\% \cref{subsec:Numerical results}.\\\\

\captionsetup[subfigure]{textfont=normalfont,singlelinecheck=off,justification=raggedright}

\subsection{Numerical simulations} \label{subsec:Numerical results}
\begin{figure}
 \centering
% \begin{subfigure}{0.49\textwidth}
% 	\centering
% 	\includegraphics[width=1.0\textwidth,trim={0cm 0.6cm 0.1cm 0.0cm},clip]{figures/TKE_F075_3.pdf}
% 	\caption{}  \label{subfig:tkeF075}
% \end{subfigure}
 %\hfill
 \begin{subfigure}{0.6\textwidth}
	\centering
	\includegraphics[width=1\textwidth,trim={0cm 0.25cm 0.1cm 0.0cm},clip]{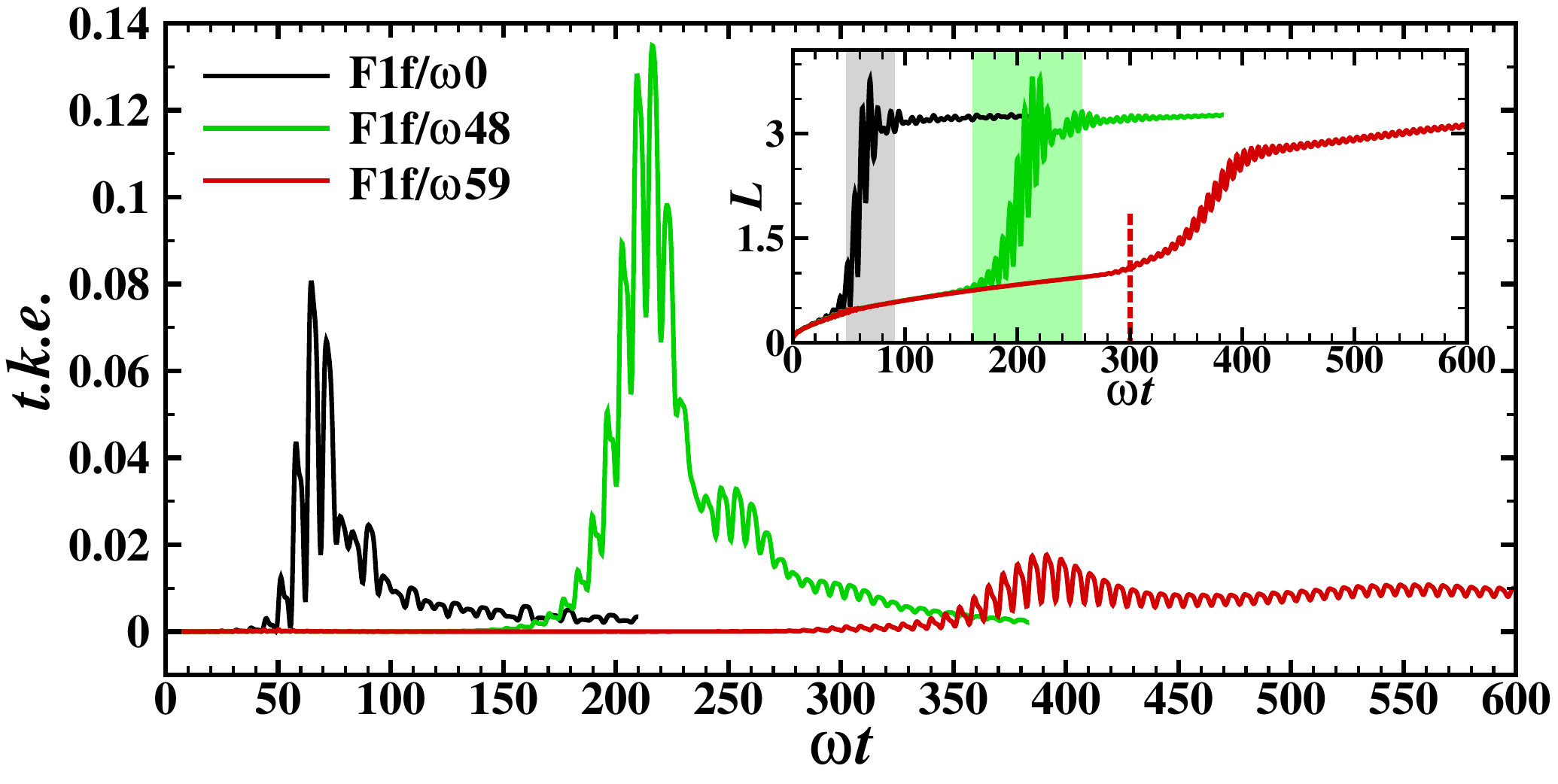}%{figures/TKE_F1_3.pdf}
	\caption{}	\label{subfig:tkeF1}
 \end{subfigure}
% \begin{subfigure}{0.49\textwidth}
% 	\centering
% 	\includegraphics[width=1.0\textwidth,trim={0cm 0.25cm 0.1cm 0.0cm},clip]{figures/TKE_F2_3.pdf}
% 	\caption{}	\label{subfig:tkeF2}
% \end{subfigure}
% \hfill
 \begin{subfigure}{0.6\textwidth}
	\centering
	\includegraphics[width=1\textwidth,trim={0cm 0.25cm 0.1cm 0.0cm},clip]{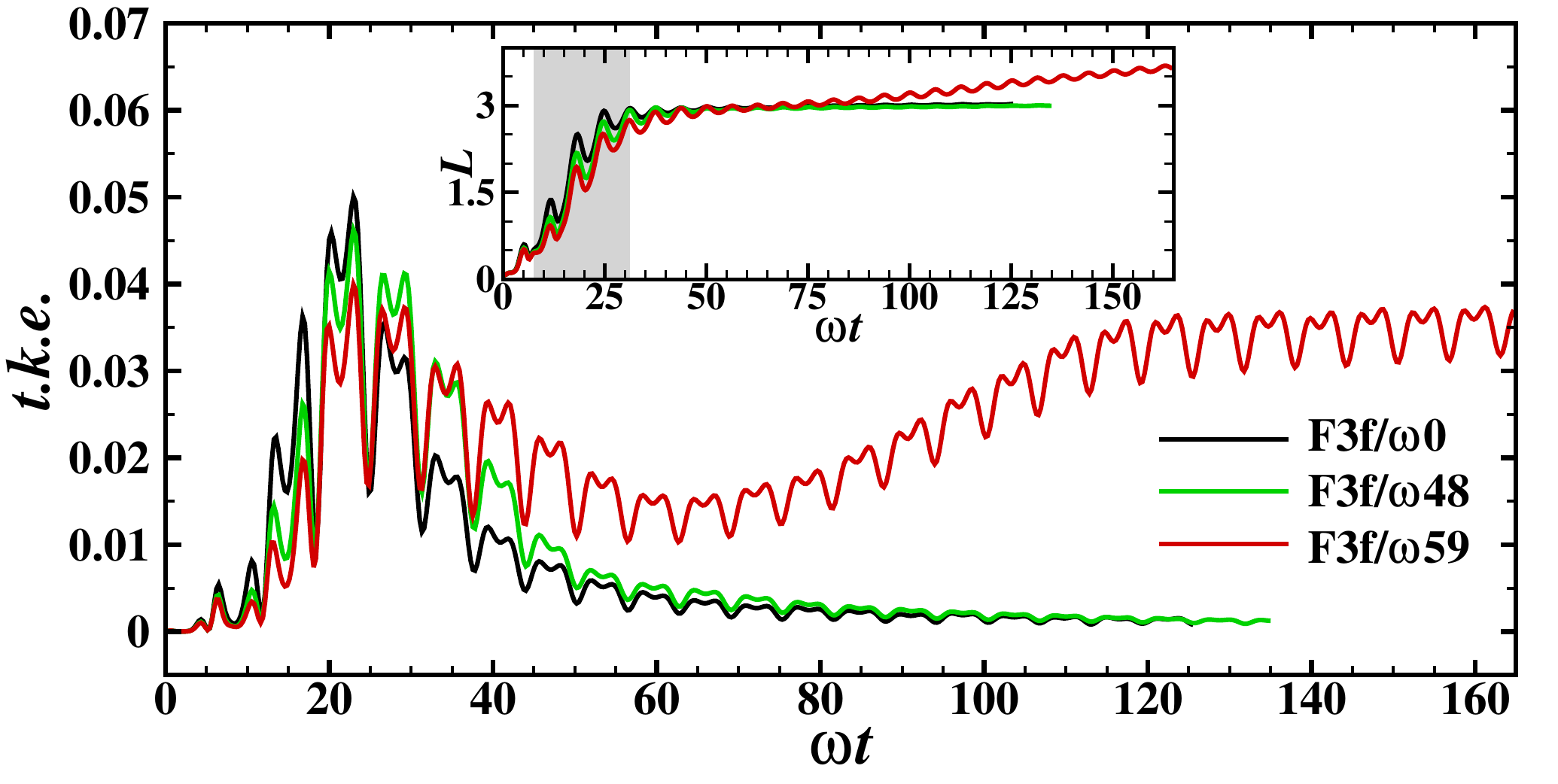}%{figures/TKE_F3_3.pdf}
	\caption{}	\label{subfig:tkeF3}
 \end{subfigure}
 \caption{Evolution of vertically ($x_3$) integrated $t.k.e.$ for both without ($f/\omega=0$) and with ($f/\omega=0.48,\,0.59$) rotation cases at (\textit{a}) $F=1$, and (\textit{b}) $F=3$. Inset shows the evolution of mixing zone size-$L$ with shaded regions indicating the sub-harmonic instability phase. The vertical red dashed line in the inset of panel (\textit{a}) denotes the onset of sub-harmonic instability for F1f/$\omega$59. } \label{fig:tke}
\end{figure}

The evolution of the vertically ($x_3$) integrated turbulent kinetic energy ($t.k.e.$) %($(\langle u_1^2 \rangle +\langle u_2^2 \rangle + \langle u_3^2 \rangle) /2$, where $\boldsymbol{u}=\boldsymbol{U}-\langle \boldsymbol{U} \rangle$),
for $F = 1$ is illustrated in figure \ref{subfig:tkeF1}. The $t.k.e.$ is defined as 
 \begin{equation}
  \label{tke}
    t.k.e. = \frac{\langle u_1^2 \rangle_H +\langle u_2^2 \rangle_H + \langle u_3^2 \rangle_H}{2}, \quad \boldsymbol{u}=\boldsymbol{U}-\langle \boldsymbol{U} \rangle_H.
 \end{equation}
Here, $\langle \boldsymbol{U} \rangle_H(x_3,t)$ is the horizontally averaged mean velocity field and $\boldsymbol{u}(x_1,x_2,x_3,t)$ is the fluctuating velocity field. The $t.k.e.$ rapidly grows for F1f/$\omega$0 and F1f/$\omega$48 during the onset of the sub-harmonic instability. Eventually, the $t.k.e.$ decays owing to the instability saturation. The $t.k.e.$ for F1f/$\omega$48 is higher than F1f/$\omega$0 due to the excitement of more unstable $\theta$-modes in the sub-harmonic region that are permitted to grow with growth rates ($\sigma$) ranging from $0$ to $0.1268$, as observed from the stability analysis. %The $t.k.e.$ decreases significantly for $f/\omega=0.59$ (case F1f/$\omega59$) owing to the generation of weaker turbulence with an increased rotation rate.
The $t.k.e.$ decreases significantly for $f/\omega=0.59$ (case F1f/$\omega59$) due to the effect of strong rotation %the generation of weaker turbulence with an increased rotation rate. 
The excitement of a small number of unstable $\theta$-modes at higher growth rates, as predicted from the linear stability analysis, is also a possible reason for the low $t.k.e.$ for F1f/$\omega59$. Although small, the $t.k.e.$ for F1f/$\omega$59 sustains owing to the continuous triggering of the sub-harmonic instability, resulting in continuous turbulent mixing. A similar evolution of $t.k.e.$ is observed for $F = 0.75$. At a higher forcing amplitude of $F = 3$, turbulence appears at the beginning, and therefore, $t.k.e.$ is similar for all $f/\omega$ cases (figure \ref{subfig:tkeF3}), indicating that the %stabilizing 
instability delaying effect of rotation is mitigated by the strong vertical forcing. Notice that the maximum value of $t.k.e.$ for F1f/$\omega$48 is higher than F3f/$\omega$48. This observation is also true for the non-rotating cases at the respective forcing amplitudes. The longer sub-harmonic instability phase for F1f/$\omega$48 is the reason for a higher value of $t.k.e.$ at F1f/$\omega$48 than at F3f/$\omega$48. The sub-harmonic instability phase %is the time interval between the onset and the saturation of the sub-harmonic instability and 
is demonstrated using the shaded regions in the evolution of the mixing zone size $L$ in the inset of figures \ref{subfig:tkeF1} and \ref{subfig:tkeF3}. A longer sub-harmonic instability phase for $F = 0.75, 1$ signifies a delay in the saturation of the instabilities triggered during the low amplitude oscillations. Therefore, the instabilities get sufficient time to evolve, eventually resulting in turbulence. In contrast to $F = 0.75, 1$, at $F = 2, 3$, the sub-harmonic instability phase is short. The instabilities saturate quickly and therefore do not get sufficient time to evolve and intensify turbulence. This observation is consistent with our linear stability analysis that predicts the maximum growth rate of the fastest growing modes is $2.7$ times greater at $F=3$ than $F=1$, resulting in the shorter sub-harmonic instability phase at $F=3$. Interestingly, both F2f/$\omega$59 (figure not shown) and F3f/$\omega$59 manifest a recovery in $t.k.e.$ owing to the continuous triggering of the sub-harmonic instabilities.\\

\captionsetup[subfigure]{textfont=normalfont,singlelinecheck=off,justification=raggedright}

\begin{figure}
 \centering
% \begin{subfigure}{0.328\textwidth}
 \begin{subfigure}{0.195\textwidth}
    \centering
     \includegraphics[width=1\textwidth,trim={0.1cm 0.1cm 0.1cm 0.0cm},clip]{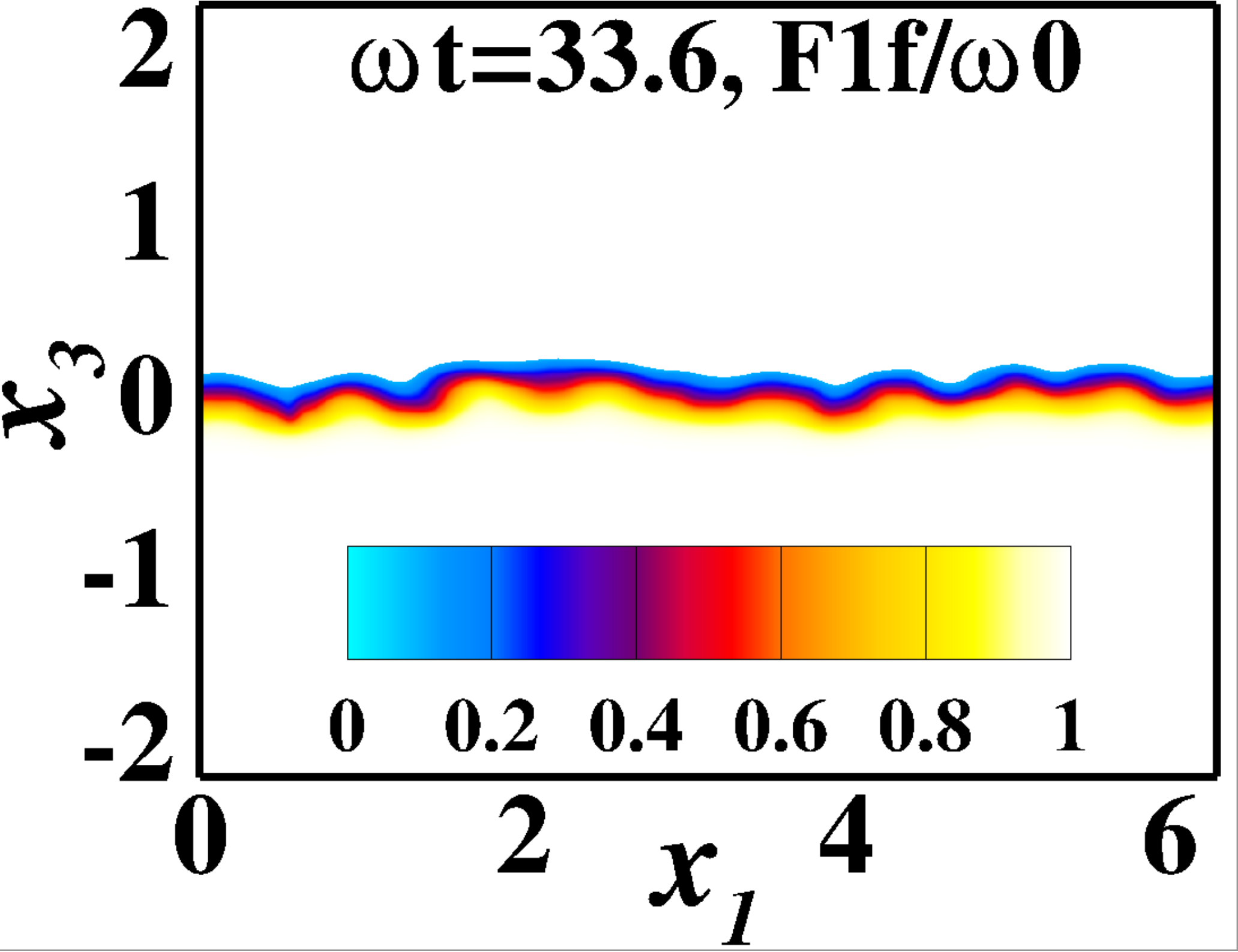}
    \caption{}  \label{subfig:C a}
 \end{subfigure}
 %\hfill
 \begin{subfigure}{0.195\textwidth}
    \centering
    \includegraphics[width=1\textwidth,trim={0.5cm 0.1cm 0.1cm 0.0cm},clip]{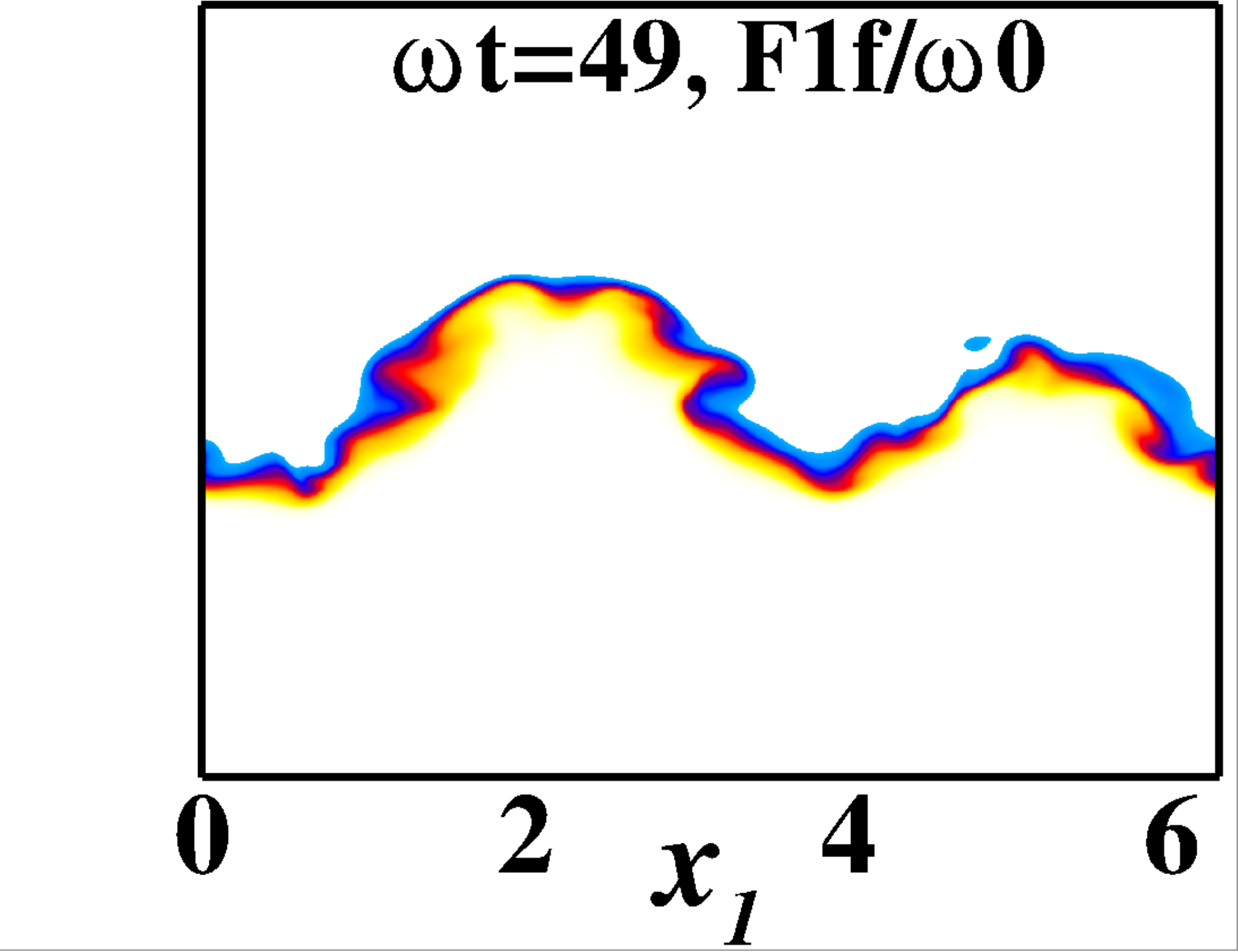}
    \caption{}  \label{subfig:C b}
 \end{subfigure}
 %\hfill
 \begin{subfigure}{0.195\textwidth}
    \centering
    \includegraphics[width=1\textwidth,trim={0.5cm 0.1cm 0.1cm 0.0cm},clip]{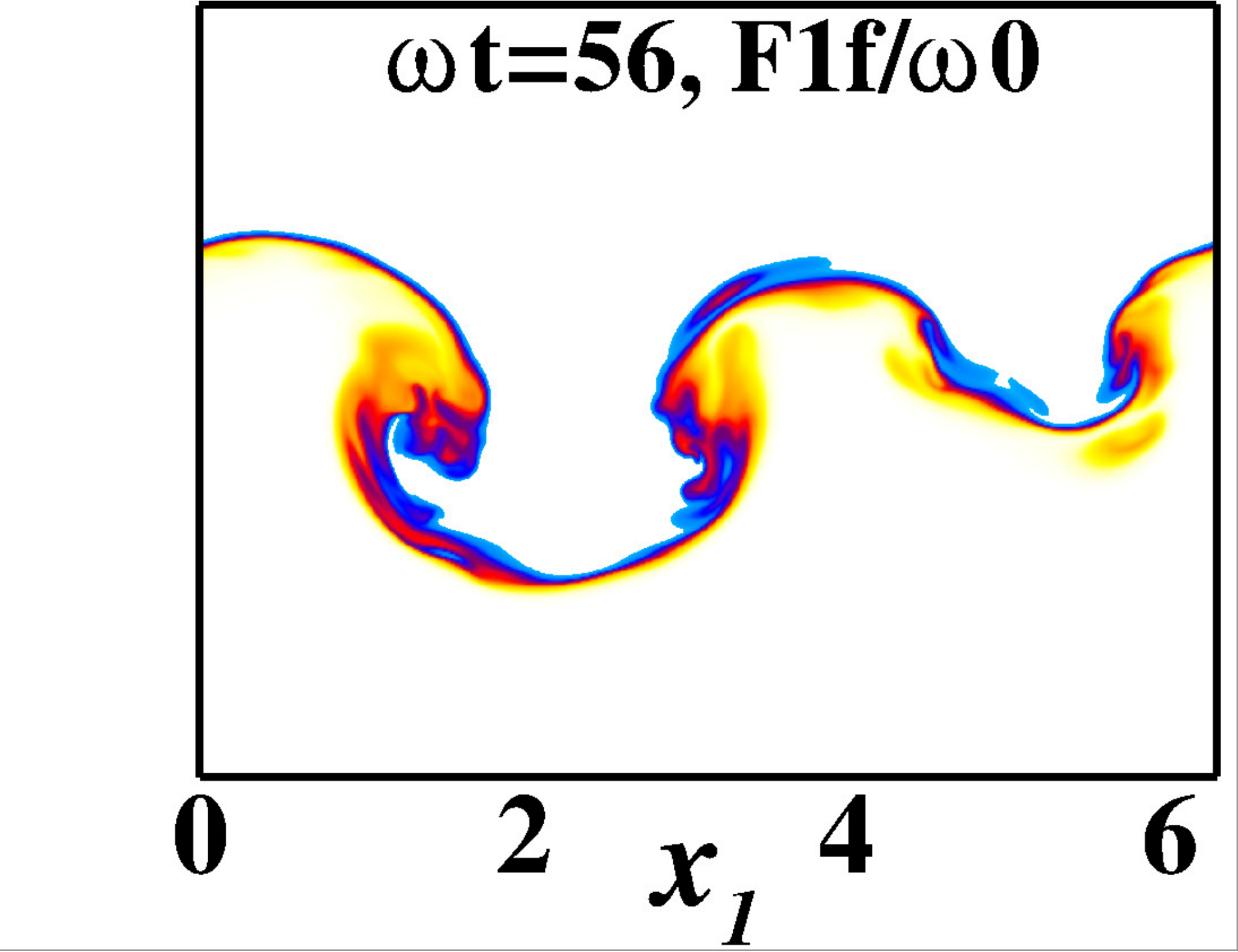}
    \caption{}  \label{subfig:C c}
 \end{subfigure}
 \begin{subfigure}{0.195\textwidth}
    \centering
    \includegraphics[width=1\textwidth,trim={0.5cm 0.1cm 0.1cm 0.0cm},clip]{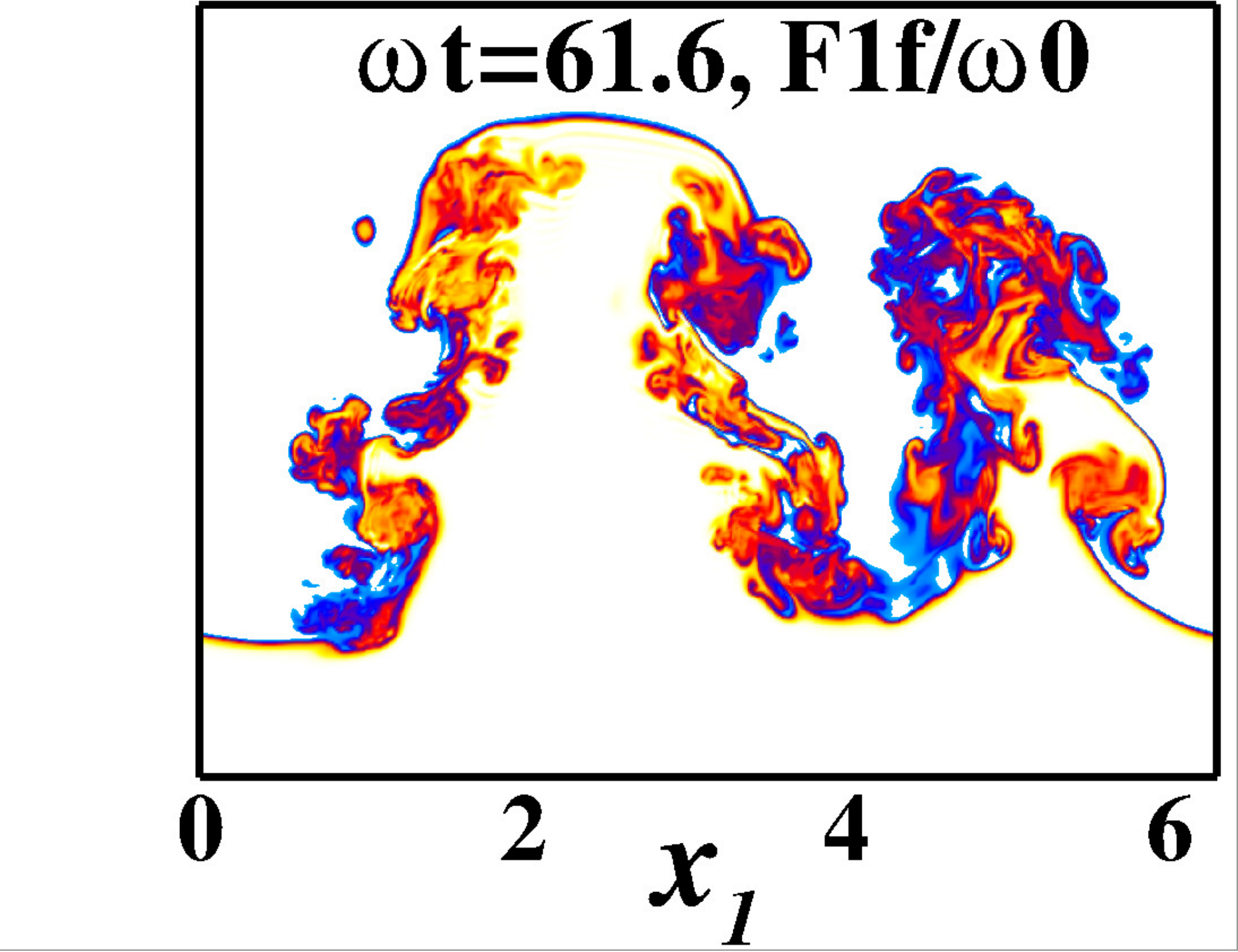}
    \caption{}  \label{subfig:C d}
 \end{subfigure}
 \begin{subfigure}{0.195\textwidth}
    \centering
    \includegraphics[width=1\textwidth,trim={0.5cm 0.1cm 0.1cm 0.0cm},clip]{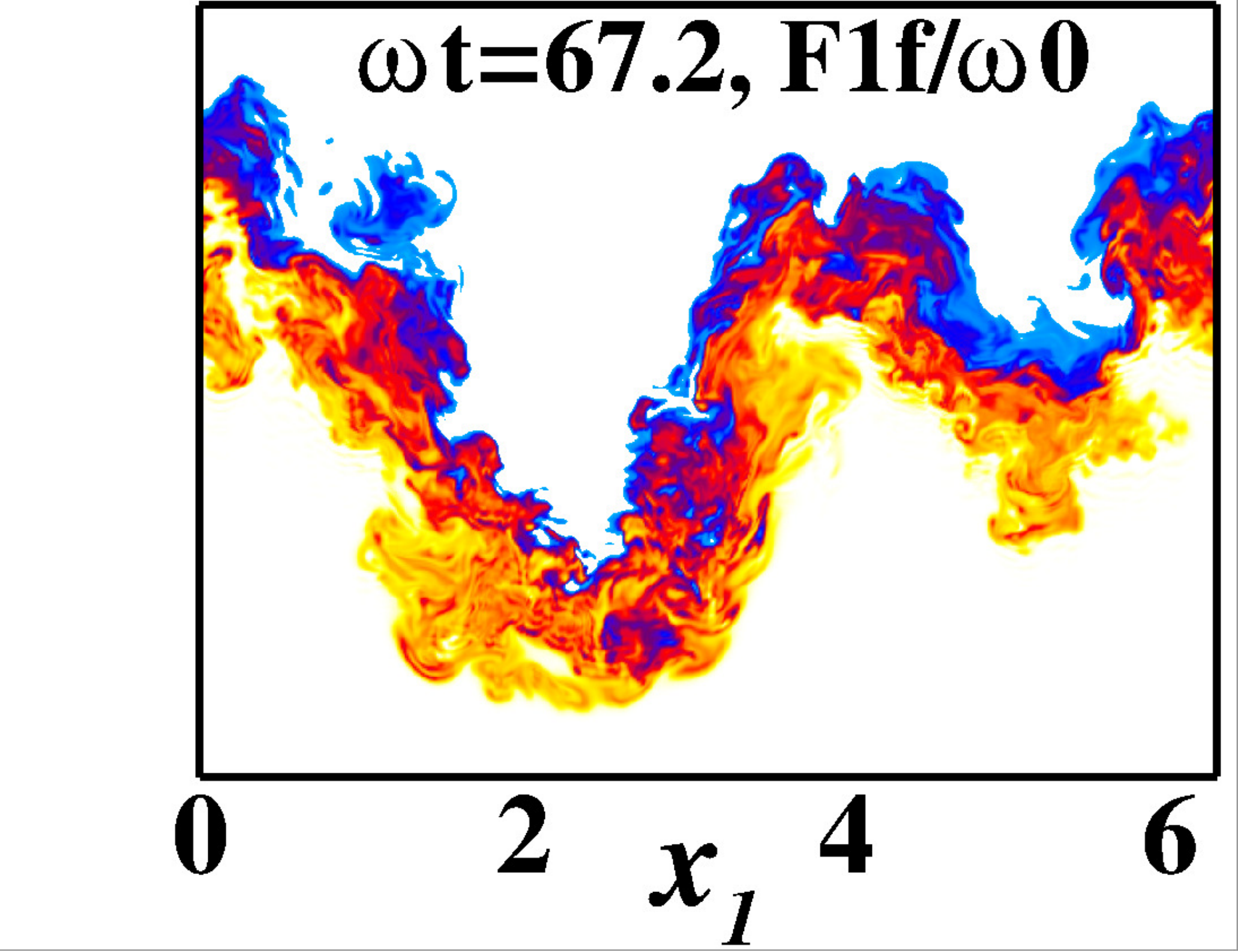}
    \caption{}  \label{subfig:C e}
 \end{subfigure}
 \caption{Contours of concentration field in vertical $x_1-x_3$ center plane ($x_2=\pi$) for case F1f/$\omega$0 at time instants (\textit{a}) $\omega t=33.6$, (\textit{b}) $\omega t=49$, (\textit{c}) $\omega t=56$, (\textit{d}) $\omega t=61.6$, and (\textit{e}) $\omega t=67.2$. Pure fluids of $C=1$ (denser) and $C=0$ (lighter) are made transparent.}   \label{fig:mean conc}
 \end{figure}

The transition to turbulence phenomenon for the cases without and with rotation is analyzed using the contour plots of the concentration fields in a vertical plane. Figure \ref{fig:mean conc} depicts the evolution of the interface of the concentration field for F1f/$\omega$0. The initially perturbed interface, subjected to vertical and periodic accelerations, deforms into a wavy form owing to harmonic and sub-harmonic resonance as shown in figure \ref{subfig:C a} at $\omega t=33.6$. This diffused wavy interface is amplified by the sub-harmonic instability, leading to the formation of mushroom-shaped waves as shown in figure \ref{subfig:C b} at $\omega t=49$. At this instance, some waves with shorter wavelengths become disorganized and interact with each other, resulting in small velocity fluctuations. In the next oscillation, the amplitude of a longer wavelength wave amplifies, leading to the birth of a well-defined mushroom-shaped wave that bursts at the nodes %owing to the Kelvin-Helmholtz (KH) instabilities 
as depicted in figure \ref{subfig:C c} at $\omega t=56$. This process is known as wave-breaking of the Faraday waves \citep{cavelier2022subcritical} and is responsible for the onset of irreversible mixing. In the successive oscillations, the entire interface breaks down into small-scale structures leading to a full transition to turbulence as manifested in figures \ref{subfig:C d} and \ref{subfig:C e}. A similar dynamics is observed for $F = 0.75$ and $f/\omega=0$ (see Movie $1$ and Movie $4$).\\

\begin{figure}
 \centering
% \begin{subfigure}{0.328\textwidth}
% \begin{subfigure}{0.246\textwidth}
%    \centering
%     \includegraphics[width=1\textwidth,trim={0.1cm 0.3cm 0.6cm 0.2cm},clip]{figures/C_F075fw48.pdf}
%    \caption{}  \label{subfig:C F075fw48}
% \end{subfigure}
 %\hfill
 \begin{subfigure}{0.325\textwidth}
    \centering
    \includegraphics[width=1\textwidth,trim={0.1cm 0.3cm 0.6cm 0.2cm},clip]{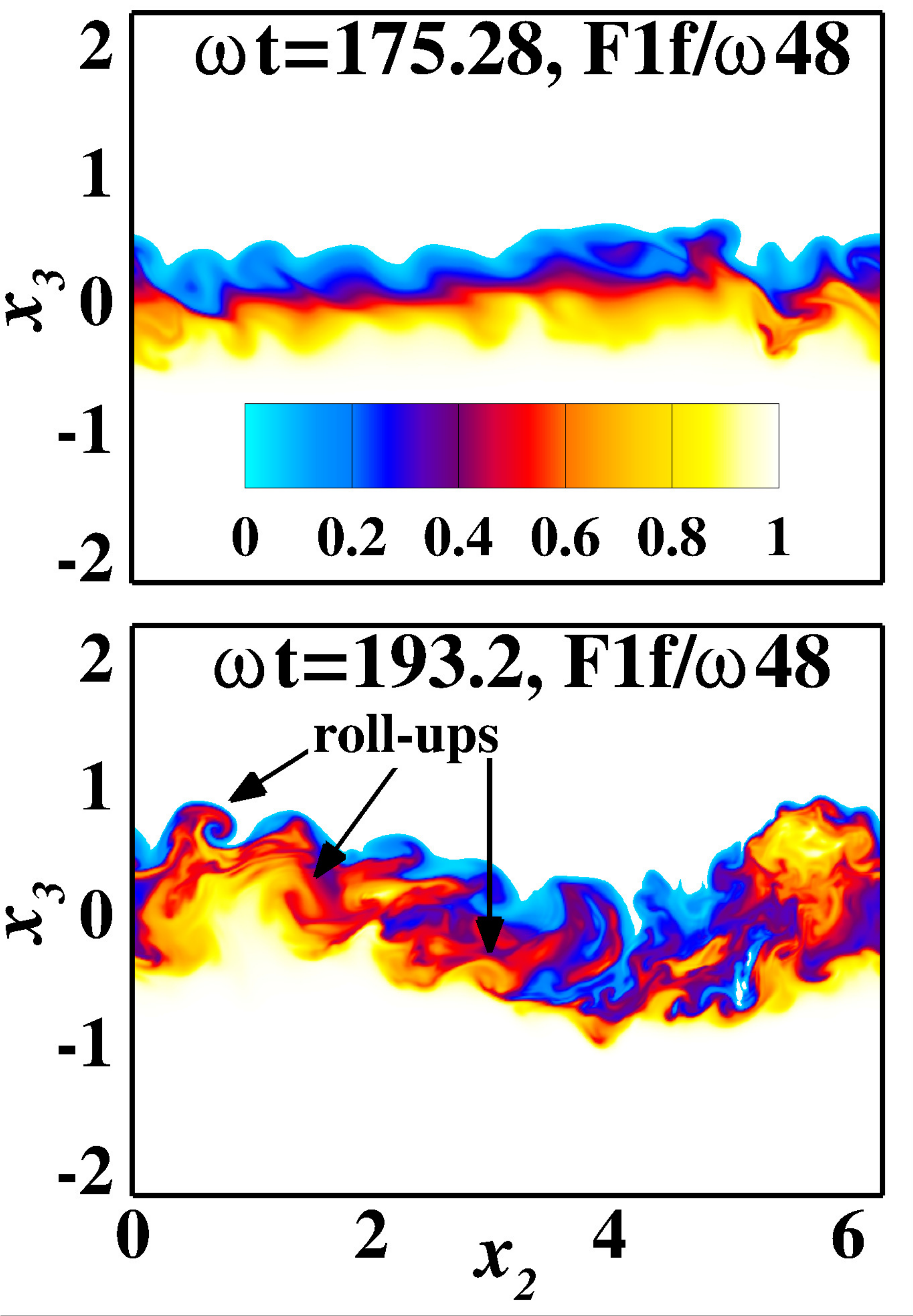}
    \caption{}  \label{subfig:C F1fw48}
 \end{subfigure}
 %\hfill
 \begin{subfigure}{0.325\textwidth}
    \centering
    \includegraphics[width=1\textwidth,trim={0.1cm 0.3cm 0.6cm 0.2cm},clip]{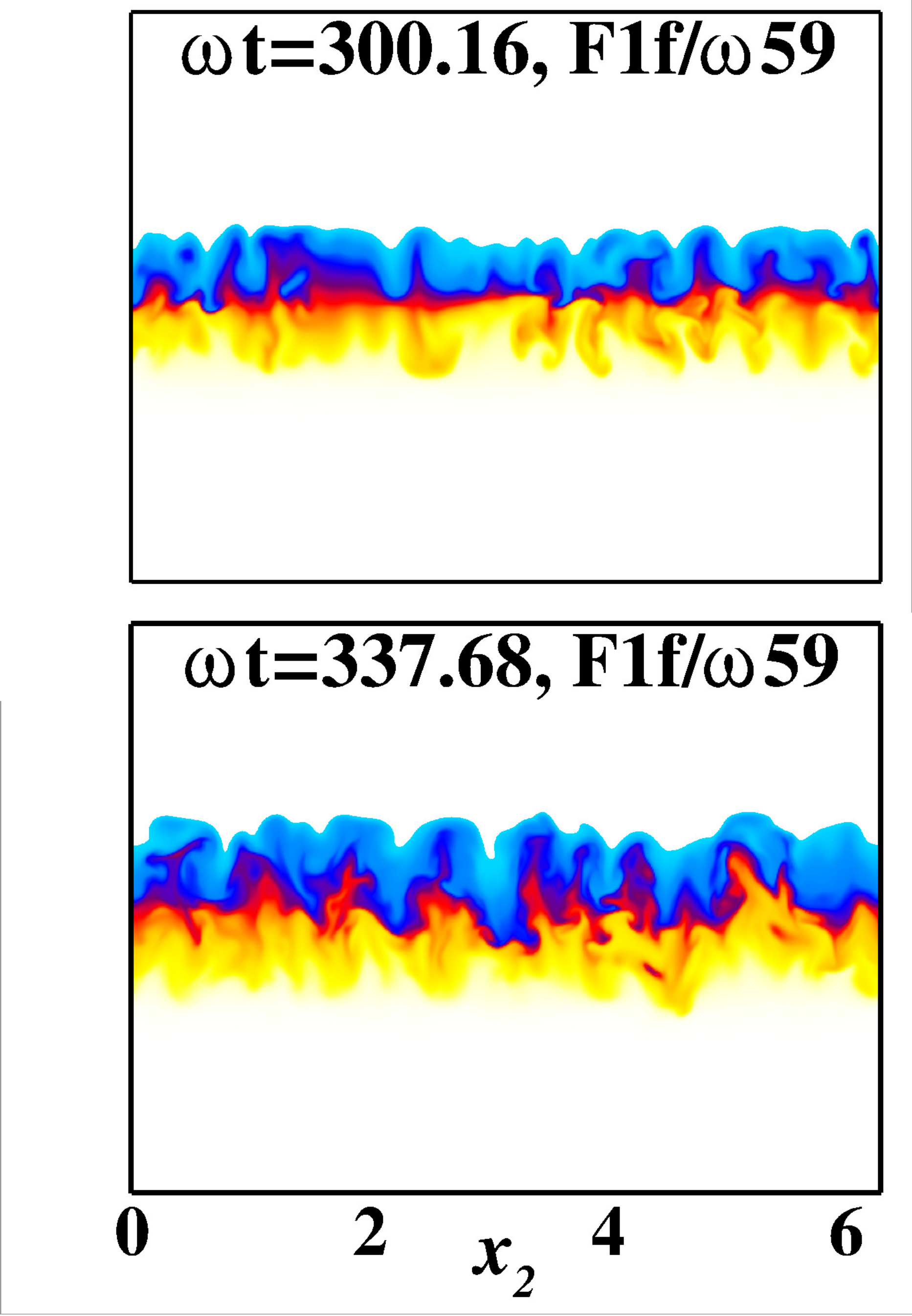}
    \caption{}  \label{subfig:C F1fw59}
 \end{subfigure}
 \begin{subfigure}{0.325\textwidth}
    \centering
    \includegraphics[width=1\textwidth,trim={0.1cm 0.3cm 0.6cm 0.2cm},clip]{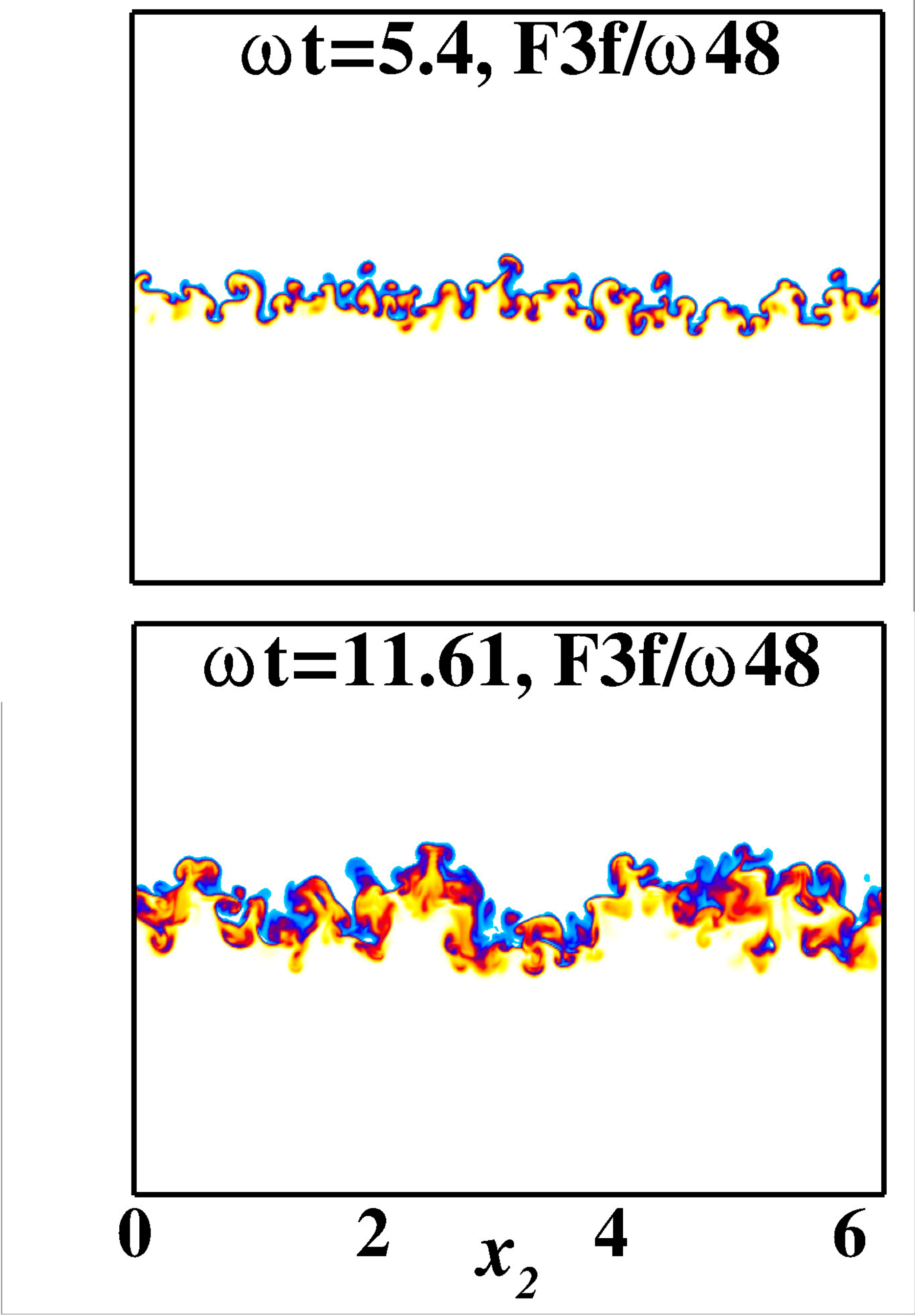}
    \caption{}  \label{subfig:C F3fw48}
 \end{subfigure}
 \caption{Contours of concentration field in vertical $x_2-x_3$ center plane ($x_1=\pi$) for rotation cases (\textit{a}) F1f/$\omega$48, (\textit{b}) F1f/$\omega$59, and (\textit{c}) F3f/$\omega$48 at different time instants during the instability growth stage. %KH billows are indicated by the arrows.
 }    \label{fig:conc}
 \end{figure}

The combined effect of vertical oscillation and rotation in F1f/$\omega48$ results in roll-ups at multiple locations on the diffuse interface as shown in figure \ref{subfig:C F1fw48} at $\omega t=175.28$. In subsequent oscillations, roll-ups will continue to form with a large size but in the opposite direction to that of the previous one. These roll-ups will finally %take the form of Kelvin-Helmholtz (KH) billows that will 
break into turbulence (see figure \ref{subfig:C F1fw48}). % at $\omega t=193.2$). 
Notice that the entire interface layer breaks into small structures in one oscillation resulting in a larger thickness of the turbulent mixing region as compared to F1f/$\omega0$ where the mushroom-shaped wave first breaks at the nodes and then turbulence spreads to the entire interface layer in the subsequent oscillations. Movie $2$ included as the supplementary data demonstrates this event. The formation of the roll-ups and their eventual breaking to turbulence is responsible for larger $t.k.e.$ in F1f/$\omega48$ as compared to F1f/$\omega0$ (figure \ref{subfig:tkeF1}). This phenomenon is consistent with our linear stability analysis as follows. For F1f/$\omega48$ a significant number of unstable modes are permitted to grow at all growth rates. This has been demonstrated in figures \ref{subfig:F1_N2W2_04} and \ref{subfig:F1_N2W2_03} when $\left(N/\omega \right)^2=0.4$ and $0.3$ respectively. The $\omega t$ corresponding to $\left(N/\omega \right)^2=0.4$ and $0.3$ is in the range $175-205$. Figure \ref{subfig:C F1fw48} shows the formation of the roll-ups at $\omega t=193.2$, and therefore it can be deduced that the unstable modes promote the formation of roll-ups that break to enhance $t.k.e$ for F1f/$\omega48$.  We observe similar dynamics for F075f/$\omega48$ (see Movie $5$). This phenomenon results in a gradual increase of irreversible mixing. When the Coriolis frequency increase to $f/\omega=0.59$ (case F1f/$\omega59$), we found that the effect of rotation is strong enough to suppress the formation of large amplitude waves and roll-ups. Therefore, the turbulent mixing is only caused by the finger-shaped instability %{\color{red}Can we say that these are Rayleigh-Taylor instabilities?} {\color{blue}``I do not think that these are Rayleigh-Taylor instabilities, because in Rayleigh-Taylor instability, the denser fluid which is placed above the lighter fluid, tries to move downward due to gravity resulting in the fingers of the denser fluid. And, the lighter fluid moves upward. In our case the fluids are stably stratified and the fingers are caused by the vertical oscillations, unlike the Rayleigh-Taylor instability where only gravity is responsible for the fingers.''} 
as depicted in figure \ref{subfig:C F1fw59} (see supplementary Movie $3$ and Movie $6$).\\

\captionsetup[subfigure]{textfont=normalfont,singlelinecheck=off,justification=raggedright}
\captionsetup[subfigure]{textfont=normalfont,singlelinecheck=off,justification=raggedright}
 
At higher forcing amplitude $F=3$ for $f/\omega=0.48$, we observe the development of random finger-shaped structures on the interface that break in the first oscillation as shown in the figure \ref{subfig:C F3fw48}. Therefore, the turbulent mixing starts from the beginning of the periodic forcing. Additionally, the intensity of turbulence is low because of the shorter sub-harmonic instability phase at higher $F$. As the %stabilizing 
instability delaying effect of rotation is subdued at higher forcing amplitudes \citep{singh2022onset}, the turbulent mixing occurs by this mechanism for all $f/\omega$ at $F=2,3$ (figures not shown). We have included the animation of the concentration field for F3f/$\omega48$ as supplementary data (Movie $7$).\\

Since there is no mean velocity field involved, we start the analysis of the irreversible mixing using the $t.k.e.$ evolution equation as follows: 
\begin{equation}
\label{tke budget}
 \frac{dt.k.e.}{dt}=\mathcal{P}-S-B-\epsilon-\frac{\partial \mathcal{T}_j}{\partial x_j}.
\end{equation}
The turbulent production term $\mathcal{P}=- \langle u_i u_j \rangle_H  \frac{\partial \langle U_i \rangle_H}{ \partial x_j}$ and the transport term $\partial\mathcal{T}_j/\partial x_j$, where $\mathcal{T}_j=\langle p u_j \rangle_H - 2\nu  \bigl \langle s_{ij} u_i \bigr \rangle_H + \frac{1}{2}\langle u_i u_i u_j \rangle_H$ are negligible. The energy input from the periodic forcing is $S=2\mathcal{A}g_0 F\cos{(\omega t)} \langle u_3 c \rangle_H$. The buoyancy flux $B=2\mathcal{A}g_0 \langle u_3 c \rangle_H$ accounts for the reversible rate of exchange between $t.k.e.$ and the %total 
potential energy. The viscous dissipation $\epsilon=2\nu \bigl \langle s_{ij} s_{ij} \rangle_H$, where $s_{ij}=\left( \partial u_i / \partial x_j + \partial u_j / \partial x_i\right)/2$, acts as a sink for $t.k.e.$. We obtain an excellent closure of equation \ref{tke budget} for all the simulations signifying sufficient grid spacing in all directions to resolve all length scales.\\ 

The irreversible mixing is characterized by partitioning the total potential energy (PE) $E_p= 2 \mathcal{A}g_0\int_{V} C(\boldsymbol{x}) x_3 \mathrm{d}V$ into the sum of an available potential energy (APE) $E_a$ and background potential energy (BPE) $E_b$ \citep{winters1995available}. The BPE is the minimum potential energy a flow can have and is unavailable for extraction from the potential energy reservoir for driving macroscopic fluid motion. To estimate BPE of the flow at any instance of time, a volume element $dV$ with concentration $C$ at height $x_3$ is adiabatically relocated to $x_3^*$ such that the redistributed concentration field is statistically stable ($\mathrm{d}C(x_3^*)/\mathrm{d}x_3^* \leq0$) everywhere. We compute $E_b= 2 \mathcal{A}g_0\int_{V} C(x_3^*) x_3^* \mathrm{d}V$ and subtract it from $E_p$ to obtain $E_a$ \citep{winters2013available,chalamalla2015mixing}. $E_a$ is available to be converted into kinetic energy. Since our domain is closed by the top and bottom boundaries, the surface fluxes are zero. We derive the simplified equations for the evolution of PE, BPE, and APE following \cite{winters1995available},
 \begin{subequations}
    \begin{equation}
     \label{dTPEdt}
     \frac{dE_p}{dt}= B_V+\phi_i,
    \end{equation}
    \begin{equation}
     \label{dBPEdt}
     \frac{dE_b}{dt}= \phi_d,
    \end{equation}
    \begin{equation}
     \label{dAPEdt}
     \frac{dE_a}{dt}= B_V-\left(\phi_d - \phi_i \right).
    \end{equation}
\end{subequations}
Here $B_V$, $\phi_i$, and $\phi_d$ are integrated over the volume $V=l_{x_1} l_{x_2} (2H-2H_s)$, and $(2H-2H_s)$ is the vertical height of the domain excluding top and bottom sponge layer thickness. The $B_V=\int{B l_{x_1} l_{x_2} \mathrm{d}x_3}$ is volume integrated buoyancy flux and $\phi_i=-2 \mathcal{A}g_0 \kappa A \left(\langle C_{top} \rangle_H-\langle C_{bot} \rangle_H\right)$, where $A=l_{x_1} l_{x_2}$, the rate at which the potential energy of a statically stable concentration ($C$) distribution increases in the absence of macroscopic fluid motion through the irreversible conversion of internal to potential energy. The $\phi_i$ is a strictly non-negative quantity and signifies the increase in BPE at a constant rate $\phi_i$ owing to the laminar diffusion. $\phi_d=2 \mathcal{A}g_0 \kappa \int_{V} -\frac{\mathrm{d}x_3^*}{\mathrm{d}C} |\nabla C|^2 \mathrm{d}V$ represents the rate of change of BPE due to irreversible diapycnal flux and is a measure of the irreversible mixing. Notice that as $\phi_d \geq0$, $E_b$ always increases as a result of diapycnal mixing. $\phi_d$ is associated with the increase in BPE due to laminar diffusion and energy coming directly from the $t.k.e.$ reservoir via $B_V$ which is related to the turbulent mixing. We wish to investigate the energetics associated with the turbulent mixing caused by the sub-harmonic instability. Therefore, we define the irreversible mixing rate $\mathcal{M}$ $\left(=\phi_d-\phi_i\right)$ to isolate the increase in BPE due to purely turbulent mixing from that of laminar diffusion \citep{peltier2003mixing}. Since we have periodic forcing in the domain, we choose to calculate the cumulative mixing efficiency \citep{peltier2003mixing,briard2019harmonic} as follows:
\begin{equation}
 \label{efficiency}
    \eta_{cu}(t)=\frac{\int_{0}^{t}\mathcal{M} \;\mathrm{d}t}{\int_{0}^{t}\left(\mathcal{M} +\epsilon_V\right)\mathrm{d}t}.
 \end{equation}
 Here $\epsilon_V=\int{\epsilon l_{x_1} l_{x_2} \mathrm{d}x_3}$ is the volume integrated viscous dissipation. The cumulative mixing efficiency $\eta_{cu}$ measures the irreversible loss of $t.k.e.$ that expends in irreversible turbulent mixing in contrast to the total $t.k.e.$ loss to the irreversible turbulent mixing and viscous dissipation.\\
 
 \begin{figure*}
 \centering
 \includegraphics[width=0.4\textwidth,trim={0cm 0.1cm 0.2cm 0cm},clip]{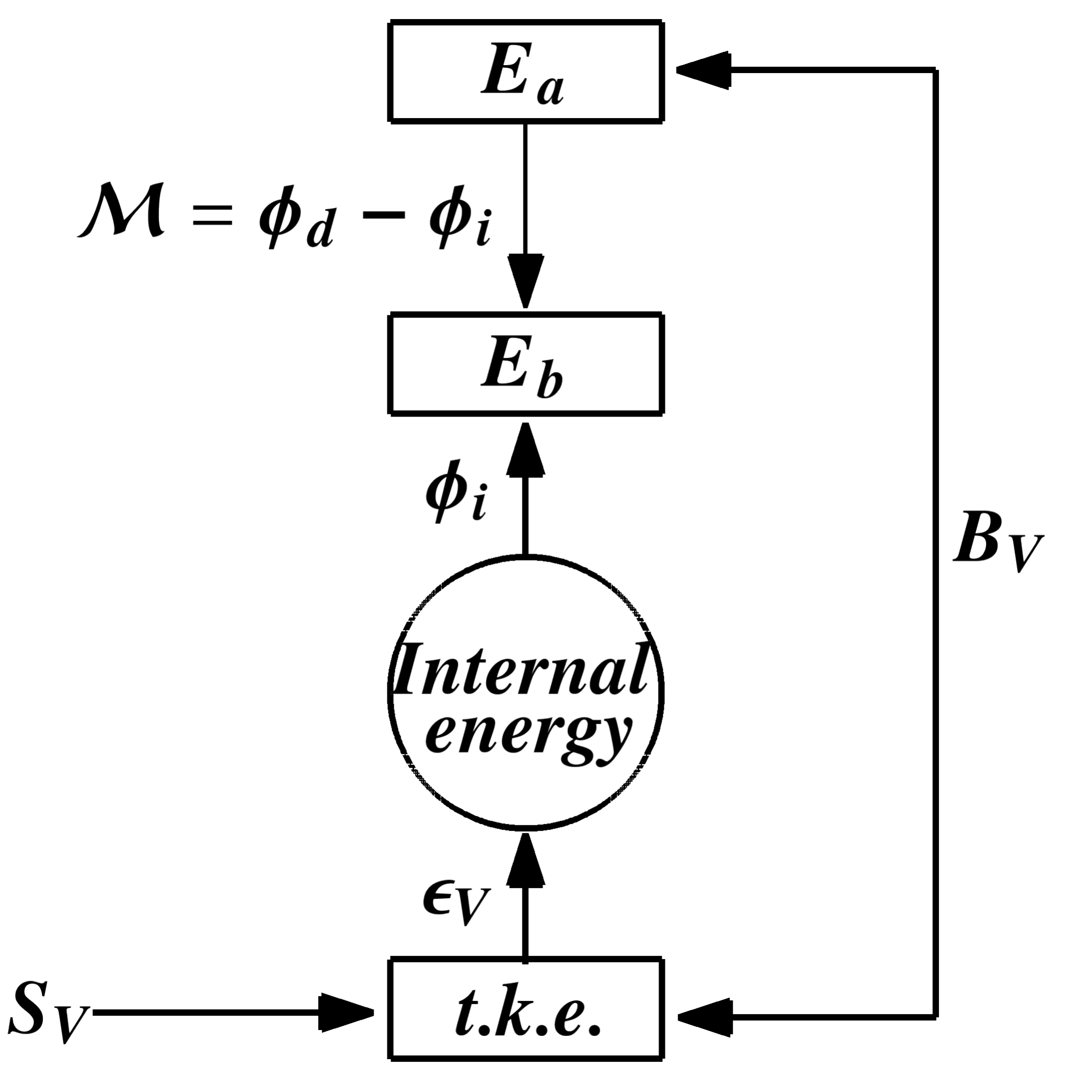}
 \caption{Schematic of energy pathways: $S_V$ is the volume integrated $S$ denoting the external input of energy from periodic forcing, $B_V$ is reversible buoyancy flux, $\mathcal{M}$ is irreversible mixing rate due to purely turbulent mixing, $\phi_d$ is irreversible diapycnal flux, $\phi_i$ is conversion of internal to potential energy due to laminar diffusion, and $\epsilon_V$ is kinetic energy dissipation. }
 \label{fig:energy pathway}
 \end{figure*}

 Figure \ref{fig:energy pathway} illustrates the energy pathways. Arrows denote the energy exchanges between different energy reservoirs. $S_V$ depicts the input rate of external energy to the $t.k.e.$ reservoir. % The potential energy is divided into two separate reservoirs of APE and BPE.
 In the presence of fluid motion due to turbulent mixing, $t.k.e.$ enters the potential energy reservoir via reversible buoyancy flux $B_V$, increasing $E_a$. Some part of $E_a$ can go to $E_b$ via purely turbulent mixing as quantified by $\mathcal{M}$, resulting in irreversible mixing. The remaining portion of $E_a$ converts back to $t.k.e.$ by $B_V$. The energy loss from the $t.k.e.$ reservoir to the internal energy reservoir occurs through viscous dissipation $\epsilon_V$. The laminar diffusion contributes energy to $E_b$ directly via $\phi_i$.\\
 %When there is no macroscopic fluid motion ($B_V=0$), the concentration profile can evolve due to molecular diffusion while maintaining its static stability. The flow always remains in its reference state of minimum potential energy, signifying $E_{PE}=E_{BPE}$ from the definition of $E_{BPE}$. Therefore, no APE is created and $\phi_d = \phi_i$ from equations \ref{dTPEdt}, \ref{dBPEdt} and \ref{dAPEdt}. This signifies that molecular diffusion will increase the BPE directly via $\phi_i$ from the internal energy of the fluid.  %It is noteworthy that the increase in $E_{BPE}$ from the APE reservoir is purely associated with turbulent mixing, and any contribution from molecular diffusion in increasing $E_{BPE}$ (via $\phi_i$) is coming from internal energy reservoir. \\
 \captionsetup[subfigure]{textfont=normalfont,singlelinecheck=off,justification=raggedright}
  \begin{figure}
 	\centering
  	\begin{subfigure}{0.48\textwidth}
  		\centering
  		\includegraphics[width=1.0\textwidth,trim={0cm 1.2cm 0.07cm 0cm},clip]{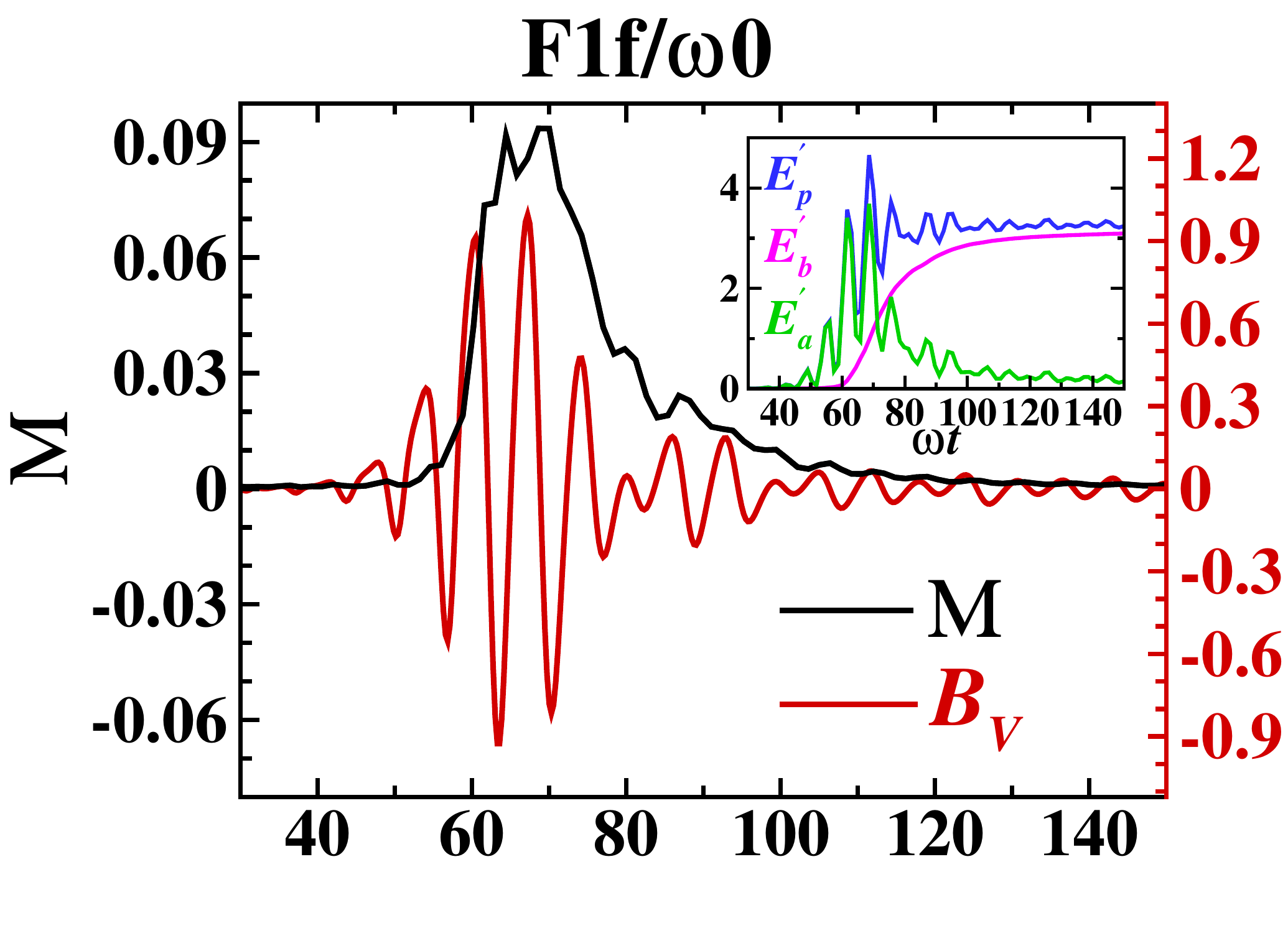}
   		\caption{}  \label{subfig:F1fw0}
  	\end{subfigure}
 	%\hfill
 	\quad 
   	\begin{subfigure}{0.48\textwidth}
   		\centering
   		\includegraphics[width=1.0\textwidth,trim={0cm 1.2cm 0.07cm 0cm},clip]{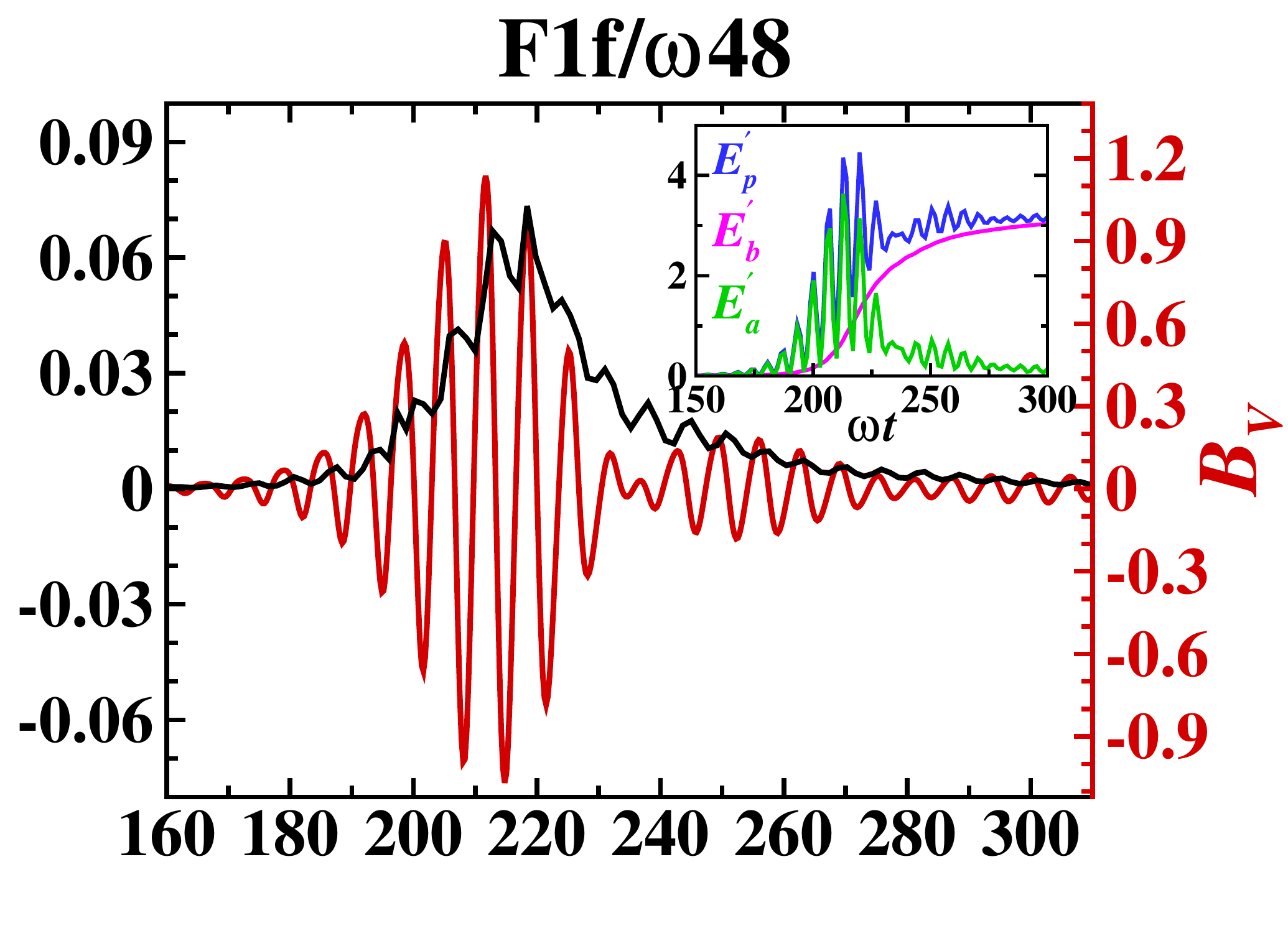}
   		\caption{}  \label{subfig:F1fw48}
   	\end{subfigure}
   	\begin{subfigure}{0.48\textwidth}
  		\centering
   		\includegraphics[width=1.0\textwidth,trim={0cm 0.4cm 0.07cm 0cm},clip]{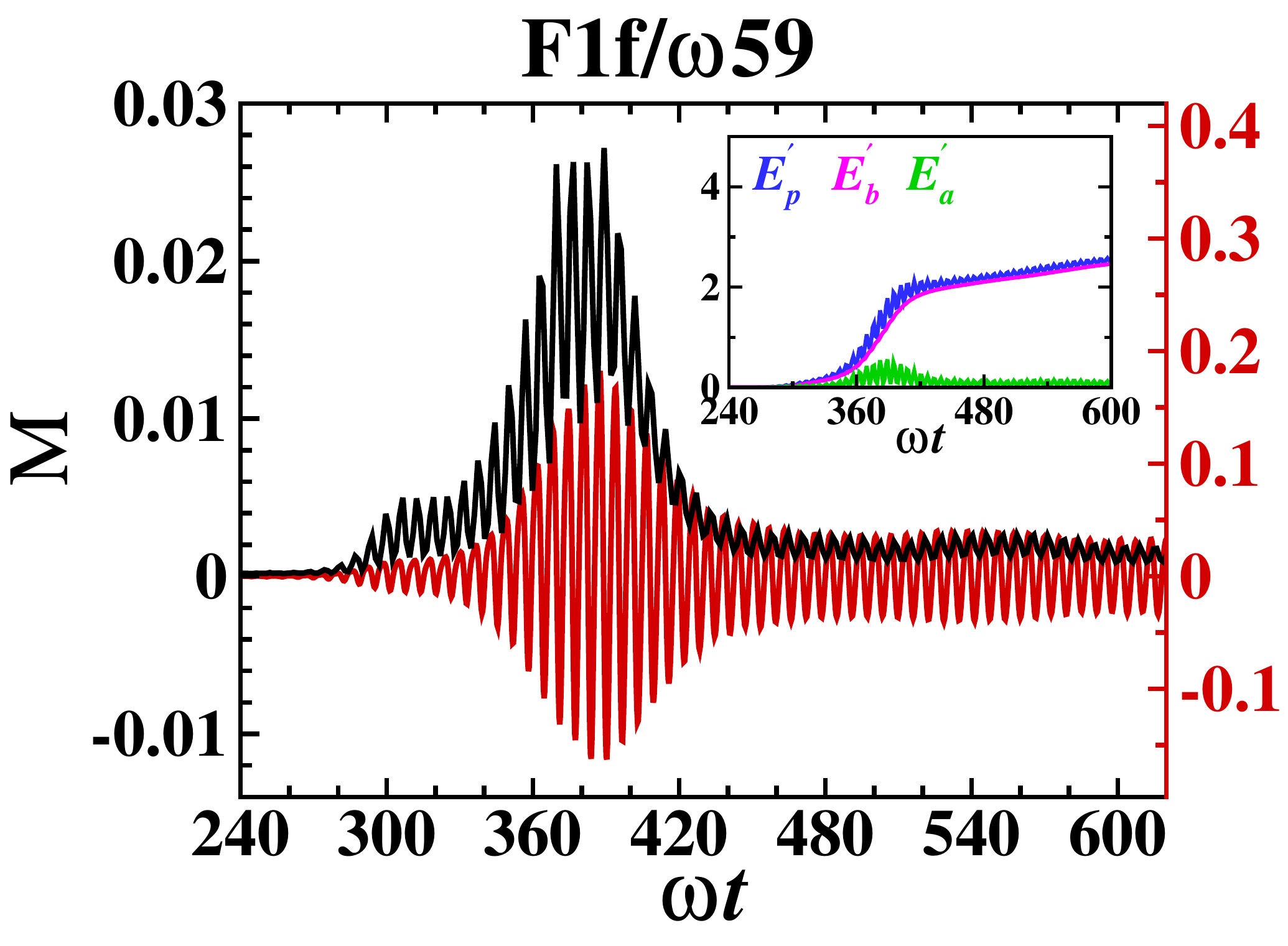}
   		\caption{}  \label{subfig:F1fw59}
   	\end{subfigure}
%   %	\begin{subfigure}{0.329\textwidth}
%  %		\centering
%  		%\includegraphics[width=1\textwidth,trim={0cm 0.4cm 0.2cm 0cm},clip]{figures/F3fw0_1.pdf}
%  	%	\caption{}  \label{subfig:F3fw0}
%  	%\end{subfigure}
%   %	\begin{subfigure}{0.329\textwidth}
%   %		\centering
%   		%\includegraphics[width=1.0\textwidth,trim={0cm 0.4cm 0.2cm 0cm},clip]{figures/F3fw48_1.pdf}
%   	%	\caption{}  \label{subfig:F3fw48}
%   	%\end{subfigure}
   	\quad
   	\begin{subfigure}{0.48\textwidth}
   		\centering
   		\includegraphics[width=1.0\textwidth,trim={0cm 0.4cm 0.07cm 0cm},clip]{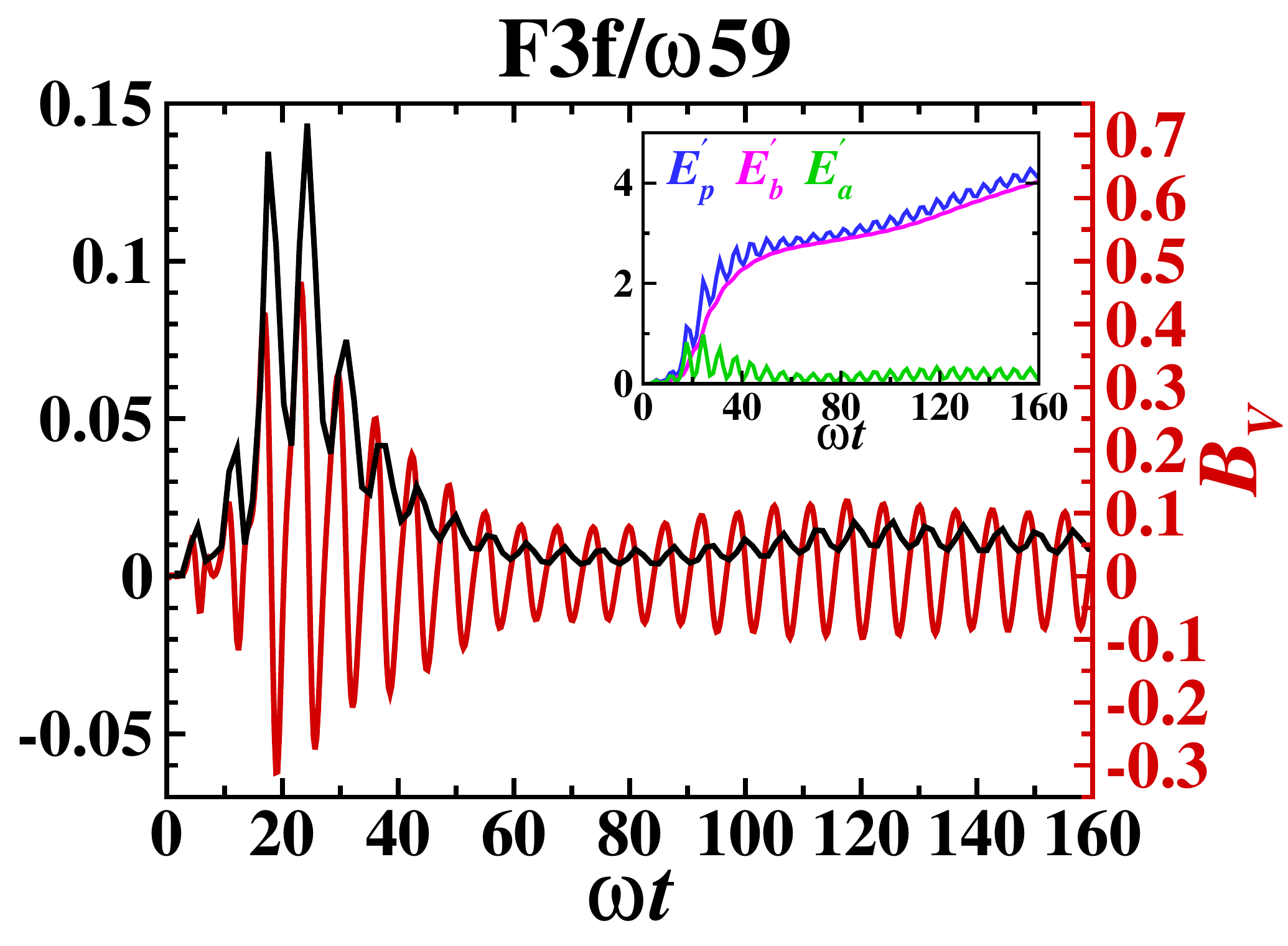}
   		\caption{}  \label{subfig:F3fw59}
   	\end{subfigure}
  	\caption{Time evolution of buoyancy flux ($B_V$) %{\color{blue}$\left(B_V=\int{B l_{x_1} l_{x_2} \mathrm{d}x_3} \right)$}, 
   and irreversible mixing rate ($\mathcal{M}$) at $F=1$ for (\textit{a}) $f/\omega=0$, (\textit{b}) $f/\omega=0.48$, (\textit{c}) $f/\omega=0.59$, and at $F=3$ for (\textit{d}) $f/\omega=0.59$. The corresponding time evolution of change in PE ($E_p'(t)=E_p(t)-E_p(0)-\phi_i t$), BPE ($E_b'(t)=E_b(t)-E_b(0)-\phi_i t$), and APE ($E_a'(t)=E_p'(t)-E_b'(t)$), due to irreversible turbulent mixing are shown in the inset of each panel.  }
%      shows the time evolution of the potential energies of the simulated,
% perturbed flow relative to their initial values. Note that the cumulative energy change
% associated with the uniform rate of laminar diffusion has been subtracted from the
% total and background potential energies, thus emphasizing the energetics associated
% with the shear instability
 \label{fig:energy rates}
\end{figure}

%   \captionsetup[subfigure]{textfont=normalfont,singlelinecheck=off,justification=raggedright}
%  \begin{figure}
% 	\centering
%  	\begin{subfigure}{0.48\textwidth}
%  		\centering
%  		\includegraphics[width=1.0\textwidth,trim={0cm 0.1cm 0.07cm 0cm},clip]{figures/tBflux_F1_F3.pdf}
%   		\caption{}  \label{subfig:F1fw0}
%  	\end{subfigure}
% 	%\hfill
% 	\quad 
%   	\begin{subfigure}{0.48\textwidth}
%   		\centering
%   		\includegraphics[width=1.0\textwidth,trim={0cm 0.1cm 0.07cm 0cm},clip]{figures/phi_d_F1_F3.pdf}
%   		\caption{}  \label{subfig:F1fw48}
%   	\end{subfigure}
%   	\begin{subfigure}{0.68\textwidth}
%  		\centering
%   		\includegraphics[width=1.0\textwidth,trim={0cm 0.4cm 0.07cm 0cm},clip]{figures/TBAPE_F1_F3_a.pdf}
%   		\caption{}  \label{subfig:F1fw59}
%   	\end{subfigure}
%   	\hfill
%   	%\quad
%   	\begin{subfigure}{0.27\textwidth}
%   		\centering
%   		\includegraphics[width=1.0\textwidth,trim={0cm 0.4cm 0.07cm 0cm},clip]{figures/TBAPE_F1_F3_b.pdf}
%   		\caption{}  \label{subfig:F3fw59}
%   	\end{subfigure}
%  	\caption{Time evolution of (\textit{a}) buoyancy flux $\left(B_V=\int{B l_{x_1} l_{x_2} \mathrm{d}x_3} \right)$, and (\textit{b}) diapycnal flux ($\phi_d$) at $F=1$ for $f/\omega=0$, $0.48$, $0.59$, at $F=3$ for $f/\omega=0.59$. Evolution of TPE, BPE, and APE (\textit{c}) at $F=1$ for $f/\omega=0$, $0.48$, $0.59$, and (\textit{d}) at $F=3$ for $f/\omega=0.59$.}    \label{fig:energy rates}
%   \end{figure}

Figure \ref{fig:energy rates} illustrates the temporal evolution of the reversible buoyancy flux $B_V$ and the irreversible mixing rate $\mathcal{M}$. We also include the evolution of $E_p'$, $E_b'$, and $E_a'$ in the inset of each figure. $E_p'$ and $E_b'$ demonstrate the total change in PE and BPE due to irreversible turbulent mixing and are calculated relative to their initial values. $E_a'$ (=$E_p'-E_b'$) denotes the corresponding change in APE. The cumulative energy changes in PE and BPE associated with the constant rate of laminar diffusion are not included in $E_p'$ and $E_b'$. Therefore, $E_p(t)'=E_p(t)-E_p(0)-\phi_i t$ and $E_b(t)'=E_b(t)-E_b(0)-\phi_i t$, where $t$ is the time from the start of the simulation. The $t.k.e.$ for F1f/$\omega$0 increases due to the onset of sub-harmonic instabilities and the formation and breaking of mushroom-shaped waves. This increased $t.k.e.$ results in an increase in $B_V$ at $\omega t\simeq46-55$, and therefore, the potential energy of the flow also increases as demonstrated in figure \ref{subfig:F1fw0}. The PE is stored as APE, which almost returns to $t.k.e.$ via $B_V$ because the transfer of APE to BPE remains zero as quantified by $\mathcal{M}\approx0$. In the subsequent oscillation, this $t.k.e.$ contributes to the external energy input, resulting in the increased $t.k.e.$ and wave-breaking. The $\mathcal{M}$ also starts increasing after $\omega t \simeq55$. After wave-breaking, the entire interface layer breaks, resulting in a rapid increase of $\mathcal{M}$ signifying rapid conversion of APE into BPE, and the APE decreases. %{\color{blue}\sout{A fraction of this PE is APE. Some part of this APE can convert back to $t.k.e.$ via $B_V$ and the rest can convert to internal energy that increases BPE owing to $\phi_i$. Since $B_V$ increases, APE also increases at $\omega t\simeq46-55$. The remaining portion of PE goes to BPE via $\phi_d$. The $\phi_d$ also starts increasing after $\omega t \simeq55$. After wave-breaking, the entire interface layer breaks, resulting in a rapid increase of $\phi_d$ signifying rapid conversion of PE into BPE}}.
When the instabilities saturate at $\omega t \simeq 70$, $t.k.e.$, and $B_V$ decrease, leading to a decrease in APE and PE saturates. The BPE also saturates, as demonstrated by a decreasing $\mathcal{M}$. Similar to F1f/$\omega$0, we observe that $B_V$ for F1f/$\omega$48 increases at $\omega t\simeq180$ owing to an increase in $t.k.e.$ (see figure \ref{subfig:F1fw48}). At the same instance, $B_V$, PE, and APE also increase. We find a gradual increase of $\mathcal{M}$ and BPE for F1f/$\omega$48 compared to a rapid increase for F1f/$\omega$0. The reason is the continuous formation and breaking of the roll-ups %KH billows
at the diffuse interface during the sub-harmonic instability phase for F1f/$\omega$48. With the saturation of the instability, $B_V$ and APE start decreasing, resulting in the increase and eventual saturation of BPE as illustrated by the gradual decrease of $\mathcal{M}$ to zero in figure \ref{subfig:F1fw48}. This gradual decrease in $\mathcal{M}$ denotes that irreversible mixing is sustained for an extended period than F1f/$\omega$0, owing to the long sub-harmonic instability phase. For F1f/$\omega$59, there is a significant delay in the onset of sub-harmonic instability due to the presence of the stable region where no harmonic or sub-harmonic modes are excited \citep{singh2022onset}. This stable region is apparent in the evolution of mixing zone size-$L$ for F1f/$\omega$59 as shown in the inset of figure \ref{subfig:tkeF1}. Therefore, $B_V$ and APE $\sim 0$ till $\omega t \simeq 280$. During this period, the initial concentration profile diffuses while maintaining its static stability \citep{winters1995available}. %, increasing the PE at the rate $\phi_i$.
The flow remains in its reference state of minimum potential energy till the onset of turbulent mixing, signifying $E_p=E_b$ from the definition of $E_b$. No APE is created and $\phi_d = \phi_i$ (from equations \ref{dTPEdt}, \ref{dBPEdt} and \ref{dAPEdt}) signifying the increase in BPE via $\phi_i$. Therefore, the potential energy increases at the expense of the internal energy of the fluid. We find, $\phi_d (0.001) \approx \phi_i (0.0008)$ till $\omega t \simeq 280$. %Since APE remains close to zero, PE $\approx$ BPE, and $\phi_d (0.001) \approx \phi_i (0.0008)$ from equation \ref{dAPEdt} till $\omega t \simeq 280$.
A similar phenomenon is also observed for F1f/$\omega$48, and for all $f/\omega$ at $F=0.75$.
%This is demonstrated in figure \ref{subfig:F1fw59}.
For higher rotation rate the turbulence is weaker %{\color{blue}\sout{suppresses turbulence}}
(see Movie $3$), resulting in smaller $t.k.e$ and $B_V$ (figure \ref{subfig:F1fw59}) compared to the previous cases. $B_V$ increases after $\omega t \simeq 280$, increasing APE. The major portion of the APE coming from $t.k.e$ via $B_V$ is transferred to BPE via $\mathcal{M}$. As a result, only a minor portion of the APE is available for reversible exchange back to $t.k.e.$, as shown by small APE in the inset of figure \ref{subfig:F1fw59}. %{\color{blue}\sout{However, smaller $B_V$ signifies that only a minor portion of the PE is available for conversion to $t.k.e.$ as shown by small APE in the inset of figure \ref{subfig:F1fw59}. The bulk of PE goes in BPE, and the evolution of $\phi_d$ shows its rate of increase.}}
Therefore, in the subsequent oscillation, only a small amount of kinetic energy is available in the $t.k.e.$ reservoir, which contributes to the external energy input, resulting in weaker turbulence. The instability never saturates for F1f/$\omega$59 and triggers continuously, resulting in the sustenance of $t.k.e.$ and $B_V$. The PE also continues to increase due to $B_V$, and APE continues to remain small since much of the APE goes to BPE. $\mathcal{M}$ oscillates and remains greater than zero indicating a continuous increase in BPE with time and signifying the sustenance of irreversible mixing for higher rotation rates. \\

For higher forcing amplitudes $F = 2, 3$ and $f/\omega = 0, 0.48$, the $t.k.e.$ and $B_V$ increase from the beginning of the periodic forcing, resulting in the increase of APE (figure not shown). Subsequently, BPE and $\mathcal{M}$ also increase. Since the instabilities saturate quickly for these cases, $B_V$ and APE become approximately zero. The BPE saturates, and the $\mathcal{M}$ also tends to zero, signifying the ceasing of irreversible mixing. The initial evolution of energetics for $F = 2,3$ for $f/\omega = 0.59$ (figure \ref{subfig:F3fw59} for F3f/$\omega$59) is similar to the cases discussed above. However, at a later stage, the evolution resembles the case F1f/$\omega$59 and demonstrates non-zero $B_V$ and $\mathcal{M}$, signifying a continuous increase in BPE and sustenance of irreversible turbulent mixing. \\%We summarize the energy pathways in figure \ref{subfig:BPE}.\\

 \begin{figure*}
 \centering
 \includegraphics[width=1.0\textwidth,trim={0cm 0.0cm 0.0cm 0cm},clip]{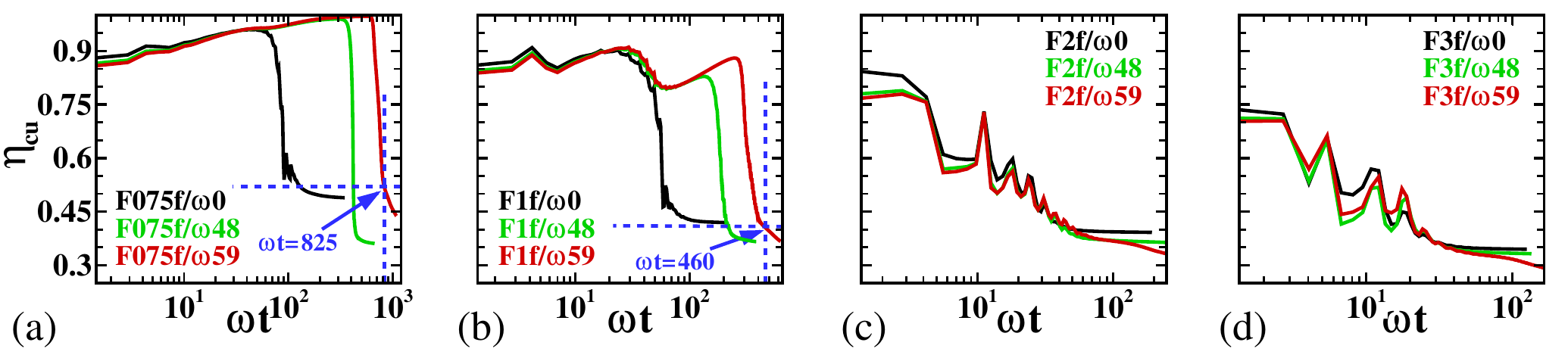}
 \caption {Time evolution of cumulative mixing efficiency $\eta_{cu}$ for each without rotation ($f/\omega=0$) and with rotation ($f/\omega=0.48,\,0.59$) cases at forcing amplitudes (\textit{a}) $F=0.75$, (\textit{b}) $F=1$, (\textit{c}) $F=2$, and (\textit{d}) $F=3$.}  \label{fig:PE eff}
 \end{figure*}

The time evolution of the cumulative mixing efficiency $\eta_{cu}$ is plotted in figure \ref{fig:PE eff}. We report the $\eta_{cu}$ values when the instability saturates for $f/\omega = 0$ and $0.48$ cases. Since the instability never saturates for $f/\omega = 0.59$, the $\eta_{cu}$ is given when $\mathcal{M}$ decreases to its approximately minimum value after reaching its peak. For F075f/$\omega$0 case, $\eta_{cu}$ decreases rapidly when the irreversible mixing rate $\mathcal{M}$ and the viscous dissipation $\epsilon_V$ increases. It attains a minimum value of $\sim 0.495$ after instability saturation. In the presence of rotation for F075f/$\omega$48, the $\eta_{cu}$ reduces to $\sim 0.363$, which implies that the mixing is not as efficient as for F075f/$\omega$0. For F075f/$\omega$59, the $\eta_{cu}$ decreases continuously owing to the continuous irreversible mixing. However, the mixing is still more efficient ($\eta_{cu}\approx 0.52$ at $\omega t\simeq825$) than F075f/$\omega$0 and F075f/$\omega$48. For F1f/$\omega$0, $\eta_{cu}$ is $\sim 0.421$ similar to that obtained by \citet{briard2019harmonic}. At $F=1$ and $f/\omega = 0.48$ $\eta_{Cu} \sim 0.372$ whereas for $f/\omega = 0.59$ the cumulative mixing efficiency is 0.406 at $\omega t = 460$, but keeps decreasing with time. We find that the mixing becomes less efficient with an increase in $F$ for all $f/\omega$ values, and the mixing efficiency continues to decrease for the $f/\omega=0.59$ cases. %Note that the cumulative mixing efficiencies obtained in the present simulations are similar to that of stratified shear turbulent flows % \citep{gregg2018mixing,caulfield2021layering,lewin2021influence}. \citep{gregg2018mixing}.
In the present work, the primary mechanism for mixing vertically oscillating stably stratified fluids is the Faraday instability. In density-stratified shear flows the Kelvin-Helmholtz instability is responsible for mixing. Despite this difference in the mixing mechanisms, the cumulative mixing efficiencies obtained in our simulations are surprisingly similar to those of the stratified shear turbulent flows \citep{gregg2018mixing}. %{\color{red}Although you mentioned the difference between our cases and shear flows in the introduction, I think you should write a few lines here as well, and mention that surprisingly the efficiency turns out to be the same. }

\section{Conclusions}\label{sec:conclusions}
We perform DNS to investigate the effect of rotation on the turbulent mixing driven by the Faraday instability in two miscible fluids subjected to vertical oscillations. At lower forcing amplitudes $F=0.75$ and $1$, the $t.k.e.$ increases with an increase in the Coriolis frequency from $f/\omega=0$ to $0.48$. This enhancement is attributed to the excitement of more unstable $\theta$-modes in the sub-harmonic region that results in %are permitted to grow with growth rates ($\sigma$) ranging from $0$ to $0.1268$ as predicted by the stability analysis. {\color{blue}These Another reason for this enhancement %This results in the
the development and breaking of roll-ups %KH billows
at the interface layer during the sub-harmonic instability phase. The mechanism of turbulence generation for the cases without rotation is different. A well-defined mushroom-shaped wave evolves for $f/\omega=0$ cases that disintegrate at the nodes. Subsequently, the turbulence spreads to the entire mixing layer in successive oscillations. In contrast to $f/\omega=0, 0.48$, $f/\omega=0.59$ have a weaker %significantly suppresses
turbulence owing to the development and gradual breaking of finger-shaped structures. At higher forcing amplitudes $F=2$ and $3$ for $f/\omega=0$, $0.48$, and $0.59$, we observe the breakdown of finger-shaped structures in the first two oscillations of periodic forcing resulting in the early onset of turbulence. However, the turbulence is less intense and short-lived than at lower forcing amplitudes owing to the shorter sub-harmonic instability phase. This shorter sub-harmonic instability phase at higher $F$ is attributed to the higher growth rate of the fastest growing modes, which is 2.7 times larger for $F=3$ than $F=1$, as determined by linear stability analysis. An interesting finding is an increase in $t.k.e.$ after an initial decrease for $f/\omega=0.59$ due to the continuous triggering of the sub-harmonic instabilities.\\ 

We also analyze the energetics associated with the flow. An increase in $t.k.e.$ for the cases with $f/\omega=0.48$ at $F=0.75,1$ implies an increase in the buoyancy flux $B_V$ and therefore, APE also increases. Some part of this APE converts to BPE via irreversible mixing rate $\mathcal{M}$ related to the turbulent mixing, whereas the rest is converted back to $t.k.e.$ via $B_V$.% , whereas the rest is converted to internal energy, increasing the BPE through $\phi_i$. The rest of the TPE goes to BPE through $\phi_d$ resulting in irreversible mixing.
When the $t.k.e.$ starts decreasing due to instability saturation, $B_V$ also decreases, signifying a decrease in APE. Therefore, BPE saturates, and $\mathcal{M}$ approaches zero. Interestingly, for $f/\omega=0.48$ the irreversible mixing sustains for a longer period than $f/\omega=0$ due to a longer sub-harmonic instability phase. The cases with a higher rotation rate of $f/\omega=0.59$ demonstrate a significant delay in the onset of sub-harmonic instability, owing to which $B_V$ and APE remain zero. The initial concentration profile diffuses during this period to increase total PE at the rate $\phi_i$. This PE goes completely to BPE at the same rate since $\phi_d$ becomes equal to $\phi_i$ according to equation \ref{dAPEdt}. After the onset of the sub-harmonic instability, the $t.k.e$, $B_V$, and APE increase. However, these values remain significantly smaller than $f/\omega=0, 0.48$ cases. In contrast to cases $f/\omega=0, 0.48$, for $f/\omega=0.59$ the bulk of the portion of APE expends to BPE via $\mathcal{M}$ resulting in a transfer of a small amount of APE back to $t.k.e$ reservoir. Therefore, in the subsequent oscillation, only a small $t.k.e.$ is available to contribute to the external energy input from periodic forcing, signifying weaker turbulence compared to $f/\omega=0, 0.48$ cases. Since the instabilities never saturate for $f/\omega=0.59$, the $t.k.e.$, $B_V$, and APE although small remain non-zero. %Therefore, TPE increases continuously and expends to BPE via $\phi_d$.
The $\mathcal{M}$ remains greater than zero, signifying a continuous increase in BPE and irreversible turbulent mixing. At $F=2,3$, $\mathcal{M}$ increases from the beginning of the vertical forcing for all $f/\omega$ cases. However, the irreversible mixing for $f/\omega = 0, 0.48$ does not sustain for long owing to the quick saturation of the sub-harmonic instability. Since the instability never saturates for $f/\omega = 0.59$, $\mathcal{M}$ demonstrates continuous oscillations and therefore, continuous irreversible mixing. Finally, the mixing phenomenon is quantified using cumulative mixing efficiency. We conclude that for rotation rates $\left(f/\omega\right)^2 > 0.25$ and at lower forcing amplitudes $F = 0.75,1$ the mixing process is more efficient because the major portion of APE expends to BPE via $\mathcal{M}$, compared to $\left(f/\omega\right)^2 < 0.25$. \\

\backsection[Supplementary data]{\label{SupMat} Supplementary movies are available at https://doi.org/**.****/jfm.***...}
\backsection[Acknowledgements]{We gratefully acknowledge Dr. Vamsi K. Challamalla for providing us the computer program for the calculation of total PE, $x_3^*$, BPE and $\phi_d$. %the support of the Science and Engineering Research Board, Government of India grant no. SERB/ME/2020318.
We also want to thank the Office of Research and Development, Indian Institute of Technology Kanpur for the financial support through grant no. IITK/ME/2019194. The support and the resources provided by PARAM Sanganak under the National Supercomputing Mission, Government of India at the Indian Institute of Technology, Kanpur are gratefully acknowledged.
}

\backsection[Declaration of interests]{The authors report no conflict of interest.}

\backsection[Author contributions]{ The authors contributed equally to analysing data and reaching conclusions, and in writing the paper.}

%\backsection[Author ORCID]{Narinder Singh,  https://orcid.org/
%0000-0002-1529-1061; Anikesh Pal,  https://orcid.org/0000-0003-2085-7231 }

\appendix
\section{Floquet Analysis}\label{appA} 
%In this appendix, we briefly discuss t
The theorems related to the Floquet theory and the steps to solve the Mathieu equation (\ref{a0 eq3}) by using these theorems are briefly discussed here. Proofs of the theorems and solution steps can be found in \citet[pp. 308-317]{jordan2007nonlinear}. 
\newtheorem{theorem}{Theorem}
\begin{theorem}[Floquet's theorem]
\label{theorem1}
Let $\frac{\mathrm{d}\vec{x}}{\mathrm{d}t}=A(t)\vec{x}$, be a first order linear differential system, where $A(t)$ is $T$-periodic matrix such that $A(t+T)=A(t), \forall \:t$. This system has at least one non-trivial solution $\vec{x}=\vec{\chi}(t)$ such that 
\begin{equation}
\label{theo 1}
 \vec{\chi}(t+T)=\mu \vec{\chi}(t),\quad \forall \:t 
\end{equation}
where $\mu$ is a \textbf{characteristic number} or \textbf{Floquet multiplier}.
\end{theorem}

If $\Phi(t)$ is a fundamental solution matrix of system $\vec{\dot x}=A(t) \vec{x}$, then $\Phi(t+T)$ is also a fundamental matrix, and there exists a non-singular matrix $C$ such that, 
\begin{subeqnarray}
\slabel{matrix C1}
\Phi(t+T)&=&C\Phi(t),\quad \forall \:t\\
\slabel{matrix C2}
C&=&\Phi^{-1}(t)\:\Phi(t+T).
\end{subeqnarray}
Here, \textbf{characteristic numbers}, $\mu'\text{s}$, are the eigenvalues of $C$. We can obtain the matrix $C=\Phi(T)$ for initial conditions at $t=0$ such that $\Phi(0)={\boldsymbol I}$, where ${\boldsymbol I}$ is the identity matrix.
\begin{theorem}
\label{theorem2}
For system $\vec{\dot x}=A(t) \vec{x}$, where $A(t)$ is $T$-periodic, with \textbf{characteristic numbers} $\mu_1,\mu_2,\dots\mu_n$, the product of the \textbf{characteristic numbers} is obtained as: 
\begin{equation}
\label{theo 2}
 \mu_1\mu_2\dots\mu_n=\mathrm{exp}\left(\int_0^T \mathrm{tr} \{A(t)\}\right).
\end{equation}
\end{theorem}

The \textbf{characteristic exponent} ($\sigma$) of the system characterizing the growth rate of the instability is defined as $e^{\sigma T}=\mu$.
\begin{theorem}
\label{theorem3}
Let $C$ has $n$ distinct eigenvalues, $\mu_i$ and corresponding $\sigma_i$, i=1,2,...,n. Then system $\vec{\dot x}=A(t) \vec{x}$ has $n$ linearly independent solutions of the form:
\begin{equation}
\label{theo 3}
\vec{\chi_i}(t)=e^{\sigma_i t}\vec{P_i}(t)
\end{equation}
where $\vec{P_i}(t)$ are the periodic vector functions with period $T$.
\end{theorem}
Clearly, \textbf{characteristic exponents} $e^{\sigma_i t}$, will determine the behaviour of the solutions (\ref{theo 3}).
Now, we apply above theorems to the mathieu equation (\ref{a0 eq3}). The equation (\ref{a0 eq3}) can be rewrite as: 
\begin{equation}
     \label{mathieu 1}
     \Ddot{a}+\left( \gamma +\alpha + \beta \cos{\tau} \right)a=0,
 \end{equation}
 where, 
\begin{equation}
     \label{mathieu var}
     \gamma = \frac{f^2 \cos^2{(\theta)}}{\omega^2};\quad \alpha = \frac{N^2 \sin^2{(\theta)}}{\omega^2};\quad \beta = F\frac{N^2 \sin^2{(\theta)}}{\omega^2}.
\end{equation}
We define $a=X$ and $\Dot{a}=Y$ to express equation (\ref{mathieu 1}) as a first-order system, given as
\begin{equation}
\setlength{\arraycolsep}{5pt}
\renewcommand{\arraystretch}{1.3}
\left[
\begin{array}{ccccc}
  \Dot{X}    \\
  \displaystyle
    \Dot{Y}   \\
 \end{array}  \right] = \left[
\begin{array}{ccccc}
  0 & 1   \\
  \displaystyle
    -(\gamma+\alpha+\beta\cos{\tau})  &  0   \\
 \end{array}  \right] \left[
\begin{array}{ccccc}
  X    \\
  \displaystyle
    Y   \\
 \end{array}  \right].
\label{system}
\end{equation}
In the notation of Theorem \ref{theorem1},
\begin{equation}
\setlength{\arraycolsep}{5pt}
\renewcommand{\arraystretch}{1.3}
A(\tau) = \left[
\begin{array}{ccccc}
  0 & 1   \\
  \displaystyle
    -(\gamma+\alpha+\beta\cos{\tau})  &  0   \\
 \end{array}  \right] .
\label{matrix A}
\end{equation}
Here, $A(\tau)$ is periodic with period $2\pi$. As $\mathrm{tr}\{A(\tau)\}=0$, the product of the \textbf{Characteristic numbers} of $A(\tau)$ is $\mu_1\mu_2=\mathrm{e}^0=1$.
% \begin{equation}
% \label{product mu}
% \mu_1\mu_2=\mathrm{e}^0=1 .
% \end{equation}
For the following initial condition
\begin{equation}
\setlength{\arraycolsep}{5pt}
\renewcommand{\arraystretch}{1.3}
\Phi(0) = \left[
\begin{array}{ccccc}
  1 & 0   \\
  \displaystyle
    0  &  1   \\
 \end{array}  \right] ,
\label{matrix phi}
\end{equation}
 we obtain the matrix $C=\Phi(T)$ by numerically integrating (with MATLAB$^{\circledR}$) the system (\ref{system}) from $\tau=0$ to $\tau=2\pi$. $\mu_1$ and $\mu_2$ are the solutions of characteristic equation of $C$
\begin{equation}
\label{cha eq1}
\mu^2-(\text{sum of roots})\mu + \text{(product of roots)}=0,
\end{equation}
and using $\mu_1\mu_2=1$, we get
\begin{equation}
\label{cha eq2}
\mu^2-\phi(\gamma,\alpha,\beta)\mu + 1=0.
\end{equation}
Here, $\phi(\gamma,\alpha,\beta)$ represents the sum of roots. The solutions $\mu_1$ and $\mu_2$ of the equation (\ref{cha eq2}) are 
\begin{equation}
\label{roots}
\mu_1,\mu_2=\frac{1}{2}\left(\phi\pm\sqrt{\left(\phi^2-4\right)}\right).
\end{equation}
Therefore, different values of $\phi$ and corresponding $\mu_1$, $\mu_2$ ($\sigma_1$ and $\sigma_2$ will determine the behaviour of solutions. When $\phi>2$: $\mu_1$, $\mu_2$ are both real and positive with $\mu_1\mu_2=1$, one of them, say $\mu_1>1$ and $\mu_2<1$. The corresponding \textbf{characteristic exponents} are real and have the form $\sigma_1=\xi>0$, $\sigma_2=-\xi<0$ (because $\sigma_2=\ln({\mu_2})/T<0$). So, the general solution from Theorem \ref{theorem3} can be written as: 
\begin{equation}
\label{phi gt2}
X(\tau)=c_1e^{\xi \tau}P_1(\tau)+c_2e^{-\xi t}P_2(\tau),
\end{equation} 
 where $\xi>0$, thus $\mathrm{e}^{\xi \tau}$ corresponds to the exponential growth in time and solution becomes \textbf{unstable} with \textbf{harmonic} in nature. When $\phi<-2$: $\mu_1$, $\mu_2$ are both real and negative with $\mu_1\mu_2=1$. The form of general solution is: 
\begin{equation}
\label{phi lt2}
X(\tau)=c_1e^{\left(\xi+\frac{1}{2}\mathrm{i}\right) \tau}P_1(\tau)+c_2e^{\left(-\xi +\frac{1}{2}\mathrm{i}\right) \tau}P_2(\tau).
\end{equation} 
Here, $\mathrm{e}^{\xi \tau}$ leads to the exponential growth of solution and eventually becomes \textbf{unstable}, with \textbf{sub-harmonic} response. When $-2<\phi<+2$: $\mu_1$, $\mu_2$ are complex, we have $\sigma_1=\mathrm{i}\zeta$ and $\sigma_2=-\mathrm{i}\zeta$, where $\zeta$ is real, signifying that the
%thus we get the solution  
% \begin{equation}
% \label{phi bt 2,-2}
% X(\tau)=c_1e^{\left(\mathrm{i}\nu\right) \tau}P_1(\tau)+c_2e^{\left(-\mathrm{i}\nu\right) \tau}P_2(\tau).
% \end{equation} 
solutions are bounded at all times and oscillatory but not necessarily periodic, and lies in \textbf{stable parameter region}. When $\phi=+2$: \:$\mu_1=\mu_2=1$, one solution is periodic with period $2\pi$ and lies on the \textbf{harmonic tongue} of \textbf{stability curve} $\phi=2$.When $\phi=-2$: \:$\mu_1=\mu_2=-1$, one solution is periodic with period $4\pi$ and lies on the \textbf{sub-harmonic tongue} of \textbf{stability curve} $\phi=2$.

\bibliographystyle{jfm}
\bibliography{jfm}

%\bibliographystyle{jfm}
%\bibliography{jfm}
%Use of the above commands will create a bibliography using the .bib file. Shown below is a bibliography built from individual items.
%\begin{thebibliography}{99}

%\expandafter\ifx\csname natexlab\endcsname\relax
%\def\natexlab#1{#1}\fi
%\expandafter\ifx\csname selectlanguage\endcsname\relax
%\def\selectlanguage#1{\relax}\fi

%\bibitem[Batchelor (1971)]{Batchelor59}
%{\sc Batchelor, G.K.} 1971 {Small-scale variation of convected quantities like temperature in turbulent fluid part1, general discussion and the case of small conductivity}, {\it J. Fluid Mech.}, {\bf 5}, pp. 3-113-133.

%\end{thebibliography}

% End of file `jfm2esam.bib'.

\end{document}